\documentclass[epj]{svjour}
\usepackage{amssymb,amsmath,stmaryrd,amsfonts,amscd}
\usepackage{graphics}
\usepackage{komandi}

\begin{document}

\title{Ballistic quantum state transfer in spin chains: 
general theory for quasi-free models and arbitrary initial states}
\titlerunning{Ballistic quantum state transfer in spin chains}

\author{Leonardo Banchi\inst{1,2,3} }
\institute{
Department of Physics and Astronomy, University College
London, Gower St., London WC1E 6BT, United Kingdom
\and
Dipartimento di Fisica, Universit\`a di Firenze,
Via G. Sansone 1, I-50019 Sesto Fiorentino (FI), Italy
\and
INFN Sezione di Firenze, via G.Sansone 1,
I-50019 Sesto Fiorentino (FI), Italy}

\date{\today}

\abstract{
  Ballistic quantum-information transfer through spin chains is based on 
  the idea of making the spin dynamics ruled by collective excitations 
  with linear dispersion relation.
  Unlike perfect state transfer schemes, a ballistic transmission requires 
  only a minimal engineering of the interactions; in fact, for most practical 
  purposes, the optimization of the couplings to the ends of the chain is 
  sufficient to obtain an almost perfect transmission. 
  In this work we review different ballistic quantum-state transfer protocols 
  based on the dynamics of quasi-free spin chains, and further generalize 
  them both at zero and finite temperature. In particular, besides presenting 
  novel analytical results for XX, XY, and Ising spin models, it is shown how, 
  via a complete control on the first and last two qubits of the chain, 
  destructive thermal effects can be cancelled, leading to a high-quality 
  state transmission irrespective of the temperature.
} 
\maketitle
\section{Introduction}
\label{s.intro}

The development of novel experimental techniques capable of accessing 
and manipulating single quantum objects 
\cite{press2008complete,weitenberg2011single,chiorescu2003coherent,schmidt2003realization,vandersypen2001experimental}
has recently triggered a multifaceted interest towards the design of quantum devices and their components. 
Amongst them, a fundamental role is played by the channels used for connecting distant quantum information units, as they must supply high-quality quantum-state and entanglement transfer within the device.

When the quantum information is embodied in physical objects that one can move,
the channel can simply be a path through which the carriers go undisturbed, as optical photons do through optical fibers.
However this is not always the case, as many proposals for the realization of quantum devices are based on localized qubit, that need being connected by a further physical object, playing the role of a {\it wire}.
In this paper we focus on this latter situation and, in particular, on the case of a wire made by a one-dimensional sequence of localized quantum systems, each statically interacting with its (nearest) neighbours.
We further assume that the wire can be assembled with physical objects of the same type of those realizing the information units, so as to actually make reference to a network of qubits. In this setup, the transmission occurs thanks to the coherent collective dynamics of the components, rather than via a subsequent application of swapping gates.
This feature was indeed one of the main motivation for introducing spin chains as many-body quantum wires \cite{Bose2003a,Bose2007}.

Quantum spin chains have a complex dynamics, featuring different phenomena, ranging from the spreading of the wave-function on different sites to the scattering between elementary excitations
\cite{takahashi2005thermodynamics}. 
Consequently, if no control on these many-body phenomena is possible, a spin chain generally behaves as a very low-quality transmission channel for long distances
\cite{Bose2003a,Bayat2008a,bayat2011initializing}.
In order to increase the transfer capabilities of a spin-chain data-bus different solutions have been proposed. Some are based on the idea of engineering the bus itself, by the specific design of its internal interactions 
\cite{christandl2004perfect,kay2009review,di2008perfect}; others use the idea
of intervening on the initialization process \cite{bayat2011initializing}, 
by preparing
the bus in a configuration found to serve the purpose; and still others
use local or global dynamical control on the chain 
\cite{PhysRevA.82.052333,haselgrove2005optimal}.
 In any case, a tough external action on the physical system is required. An alternative approach is that considering a data-bus made by a uniform spin chain which is weakly coupled to the external qubits
\cite{WojcikLKGGB2005,wojcik2007multiuser,Yao2010}. 
In this setup, the transmission quality of quantum states and entanglement can be made arbitrarily high, provided the interaction between the bus and the qubits is arbitrarily small
(see also \cite{lorenzo2013quantum,paganelli2013routing}
for a related approach using strong magnetic fields). 
 This proposal theoretically leads to near-to-perfect state transfer, meanwhile being experimentally feasible.
 However, due to the weak couplings involved, the transmission times are very long, and the system can be exposed to severe decoherence effects.

In 
\cite{Leonardo10,banchi2011long,banchi2011efficient,apollaro201299,apollaro2013transport} 
a different approach has been proposed, based on the intuition
that excitations characterized by a
linear dispersion relation are substantially transmitted coherently 
along a chain. It is shown that, by properly tuning
the interactions between the information units placed at the two ends of the transmitting chain, and the chain itself, one can induce a dynamics which is ruled by excitations with linear dispersion, i.e. essentially ballistic. As a result, by means of a \emph{minimal} static optimization of the couplings, the information flows coherently through the one-dimensional wire, allowing a fast, high-quality, quantum-states and entanglement transfer, even in the limit of an infinite-length chain, i.e., in principle, over macroscopic distances. This recipe can be actually implemented in all quasi-free models:  a special emphasis is here given to spin-$\frac12$ 
XX and XY chains, but the same arguments can be applied with little effort to other many-body systems, such as
chains of (Majorana) fermions or excitonic systems.

In this paper we present a systematic analysis of minimally engineered models
for state transmission. We review some previous findings obtained for
XX and XY models 
\cite{Leonardo10,banchi2011long,banchi2011efficient,apollaro201299}
and we extend them with new analytical results and applications. 
Moreover, we introduce a two-qubit encoding protocol for achieving, 
in principle, a ballistic quantum information transfer even at room temperature.
In Section~\ref{s.quasifree} we present the necessary
formalism for evaluating the transmission quality when the dynamics is 
generated by quasi-free Hamiltonians.  We introduce 
the optimal average fidelity as a measure of the transmission quality and we 
show the relation with the optimization of a fermion quantum walk on a line.
In Section~\ref{s.xx} we review the optimal ballistic dynamics in the XX model
and
set up the necessary formalism for understanding the state transmission as a 
wave-packet travelling in one dimension. 
In Section~\ref{s.optimalxry} we extend the 
results to XY models: even if the resulting dynamics 
does not conserve the number of fermions in the initial state, we show that
the evolution can still be described, approximately, as that of a 
wave-packet whose dynamics can be made coherent with a minimal engineering.
In Section~\ref{s.encoding} we propose a protocol for transferring states
irrespective of the initial state of the spin-chain data-bus.
This method requires only the ability to perform
gates and measurements on the first two and last two qubits of the chain.
Possible experimental realizations are discussed in Section~\ref{s.exp}
and conclusions are drawn in Section~\ref{s.concl}.

\section{State transfer in quasi-free Hamiltonians}
\label{s.quasifree}

We consider state transmission in quasi-free models and, in particular,
one-dimensional spin-$\frac12$ chains whose Hamiltonian 
can be mapped into a quadratic fermionic model:
\begin{equation}\label{e.Hfermi}
  H=\sum_{n,m=1}^{N} \left[ c_n^\dagger A_{nm} c_m +\frac{1}{2} (c_n^\dagger B_{nm} c_m^\dagger- c_n B_{nm}^* c_m)\right],
\end{equation}
where $N$ is the length of the chain,
$c_n^\dagger$ and $c_n$ are the creation and annihilation 
operators on site $n$, and $\{c_n,c_m^\dagger\}=\delta_{nm}$. 
An important class of quasi-free models is defined by the XY
spin-$1/2$ chain whose Hamiltonian reads
\begin{align}
  H = \sum_{n=1}^{N-1} \left(\frac{j_n+\gamma_n}2 \sigma_n^x\sigma_{n+1}^x +
  \frac{j_n-\gamma_n}2 \sigma_n^y\sigma_{n+1}^y\right) +
  \sum_{n=1}^N h_n \sigma_n^z~,
  \label{e.xy}
\end{align}
where $\sigma^\alpha_n$ are the Pauli spin operators acting on site $n$.
The above Hamiltonian describes $N$ spins in a one-dimensional lattice; 
each spin interacts with 
a (local) magnetic field $h_n$ and is coupled with its nearest neighbours 
via an anisotropic exchange interaction in the XY plane.
The mapping between the spin Hamiltonian \eqref{e.xy} and the 
fermion Hamiltonian \eqref{e.Hfermi} is 
realized by the Jordan-Wigner transformation 
$
  c_n=\prod_{m=1}^{n-1} \left(-\sigma_m^z\right) 
  {(\sigma_n^x-i\sigma_n^y)}/2. 
$
The corresponding {\it hopping matrices} entering in Hamiltonian 
\eqref{e.Hfermi} read
\begin{align}
  A &= \begin{pmatrix}
    h_1 & j_1 \\ j_1 & h_2 &j_2 \\  
    & j_2 & h_3 &j_3 
    \\&&&\ddots\\
    &&&j_{N-2} & h_{N-1}&j_{N-1}\\
    &&&&j_{N-1}&h_N\\
  \end{pmatrix}~,
  &
  B &= \begin{pmatrix}
    0 & -\gamma_1 \\ \gamma_1 & 0 &-\gamma_2 \\  
    & \gamma_2 & 0 &-\gamma_3 
    \\&&&\ddots\\
    &&&\gamma_{N-2} & 0&-\gamma_{N-1}\\
    &&&&\gamma_{N-1}&0\\
  \end{pmatrix}
  \label{e.ABmat}~.
\end{align}
The dynamics of the chain is completely specified in the Heisenberg 
representation by the time-evolved creation and annihilation 
operators $c_n(t)=e^{+iH t}c_n e^{-iH t}$ that can be written as
\cite{bayat2011initializing}
\begin{equation}\label{ck_t}
  c_n(t)=\sum_{m=1}^{N} \left[U_{nm}(t)c_m+V_{nm}(t)c_m^\dagger\right]~,
\end{equation}
for some $N\times N$ dynamical matrices $U(t)$ and $V(t)$ which are obtained 
with a suitable transformation (see Appendix~\ref{s.quaddiag}) 
of the matrices $A$ and $B$. 
Although the Jordan-Wigner mapping is non-local,
the above equation \eqref{ck_t} 
completely describe the dynamics of the boundaries of the
spin chain. For instance, 
as $c_1(t) = \sigma_1^-(t)$ it is possible to obtain all the possible 
observables on the first site. Indeed,
$\rho_1(t) =  ({1+\vec r_1(t) \cdot \vec \sigma_1})/2$, with the
{\it Bloch vector} elements
\begin{align}
  r_1^x(t) &= \mean{c_1^\dagger(t)+c_1(t)}~, &
  r_1^y(t) &= -i\mean{c_1^\dagger(t)-c_1(t)}~,&
  r_1^z(t) &= 2\mean{c_1^\dagger(t)c_1(t)}-1~,
  \label{e.r1t}
\end{align}
where the expectation values are taken over the initial ($t=0$) state
$\ket{\Psi_{\rm in}}$.
The state of the opposite spin 
$\rho_N(t) = ({1+\vec r_N(t) \cdot \vec \sigma_N})/2$, can be obtained in
a similar way: 
the Wigner-string appearing in the mapping between $c_N(t)$ and $\sigma_N^-(t)$
can be written in terms of the parity 
$\Pi=\exp\left(i\pi\sum_{n=1}^{N} c_n^{\dagger}c_n\right)\equiv 
\prod_{n=1}^{N} \left(-\sigma_n^z\right)$
which is a constant of motion. Thereby one obtains 
$\sigma_N^-(t) = \Pi c_N(t)$ and 
\begin{align}
  r_N^x(t) &= \mean{c_N^\dagger(t)\Pi+\Pi c_N(t)}~, &
  r_N^y(t) &= -i\mean{c_N^\dagger(t)\Pi-\Pi c_N(t)}~,&
  r_N^z(t) &= 2\mean{c_N^\dagger(t)c_N(t)}-1~.
  \label{e.rNt}
\end{align}

We focus on the quantum state transmission from 
the first qubit, sitting at site 1, to the one sitting at the opposite end
of the chain, i.e. at site $N$. Hereafter these two qubits are referred to
as the {\it sender} and {\it receiver}. Moreover, we call $\tilde\Gamma$ 
the chain composed by the spins 
localized in positions $2, \dots, N$. 
The internal chain acts as a noisy
quantum channel which, in general, may alter, disperse or localize the state
to be transferred. 
This noisy quantum channel can be described by a
mapping $\rho_1(0) \to \rho_N(t)$. 
If the sender is not initially entangled with 
the rest of the system, namely if
$\ket{\Psi_{\rm in}} = \ket{\Psi_1(0)}\otimes\ket{\Psi_{\tilde\Gamma}(0)}$,
then the mapping takes the form of a simple linear relation 
between the initial and final Bloch vectors
\begin{align}
  \vec r_N(t) = D(t)\,\vec r_1(0) + \vec d(t)~.
  \label{e.affine}
\end{align}
The latter equation describes a qubit-to-qubit quantum channel
\cite{nielsen2000qca}, 
provided that $D(t)$ and $d(t)$ satisfy some constraints 
\cite{fujiwara1999one,horodecki1996information}.
In quasi-free quantum channels the $3\times3$
matrix $D(t)$ only depends on $U_{N1}(t)$ and $V_{N1}(t)$, while
$\vec d(t)$ depends on $U(t)$, $V(t)$ and $\ket{\Psi_{\tilde\Gamma}(0)}$
\cite{bayat2011initializing}.
The matrix elements $U_{nm}(t)$ and $V_{nm}(t)$  
describe the probability amplitude that
a particle or a hole goes from site $m$ to site $n$ after a time $t$. 
Due to the relation between fermionic operators and spin ones at the 
boundaries, we show in the following that the transmission quality depends 
only on $u_1(t) = |U_{N1}(t)|$ and $v_1(t) = |V_{N1}(t)|$, 
provided that the initial state of the chain has a definite parity.

As a figure of merit for the transfer quality we use the optimal average
transmission fidelity which is a measures of the fidelity between
the initial state and the transmitted one. 
Before introducing this quantity, let us comment on a different 
strategy for quantum information transfer based on the
teleportation protocol. Quantum teleportation allows the state 
transmission from the sender to the receiver 
not by directly injecting particles through quantum
channels, but rather using local operations, classical communication, 
and an initially shared entangled state. In a spin chain setup, 
remote entanglement generation can be obtained by first creating entanglement
locally, say between the spin 1 and a near auxiliary spin 0, and then sending
``half of the state'' (the state of spin 1) to the remote part
(the spin $N$). If the ``half-state'' 
is perfectly
transferred, then the result is a maximally entangled state between 0 and $N$ at
the transmission time, that is remote entanglement. 
However, the spin chain acts generally 
as a noisy quantum channel, and the resulting state is
not maximally entangled. When the two remote parts share an entangled
state, which is not maximally entangled, the fidelity of teleportation is
not perfect \cite{bowen2001teleport,horodecki1999general}. 
One can show \cite{horodecki1999general,Bayat2008a} that 
the average fidelity of state transmission is equivalent to
the fidelity of the above protocol consisting of two steps: 
entanglement sharing and quantum teleportation.
Owing to this equivalence, we only study quantum state
transmission without considering the problem of entanglement transmission.

Quantifying the ability of a quantum channel to reliably transfer quantum
information is a subject of active research \cite{kretschmann2004tema}. 
Most figures of merit require the solution of complicated variational 
problems and are in general difficult to evaluate. Here 
we focus on a simpler quantity which does not measure the worst case scenario
but rather the average transmission quality. 
An average state transmission fidelity can be defined as
$\int d\psi_1 \bra{\psi_1}\rho_N(t)\ket{\psi_1}$, 
and measures 
the fidelity between the initial state $\ket{\psi_1}$ of the sender and
the (evolved) state of the receiver.
If a perfect transfer is obtained for every initial
state at a certain time $t^*$ then 
$\int d\psi_1 \bra{\psi_1}\rho_N(t^*)\ket{\psi_1}=1$. However, in a spin chain
setup, the external magnetic field may rotate the state $\ket{\psi_1}$ during
the transmission. This effect is independent on the initial state
and it is easily estimated and corrected via a 
local counter-rotation $R$. For this reason, in \cite{bayat2011initializing}
the optimal transmission fidelity 
\begin{align}
  F(t)=\max_{R\in{\rm SU}(2)} \int d\psi_1
  \bra{\psi_1}R^\dagger\rho_N(t)R\ket{\psi_1} = 
  \frac12+\frac{\delta_1(t)+\delta_2(t)+\sign[\det D(t)]
  \delta_3(t)}6 ~,
  \label{e.fidelityT}
\end{align}
has been introduced as a figure of merit for the transmission quality. 
In the above expression, $\delta_i(t)$ are 
the singular values of $D(t)$ sorted in decreasing order.
The optimal counter-rotation $R$ which maximizes the average fidelity
cancels the effects of the spin precession and it is
independent on the particular initial state: it depends only on the model 
Hamiltonian, and as such can be estimated {\it a priori}.

For quasi-free one-dimensional spin models one obtains
\begin{align}
F(t) = 
  \frac12+\frac16\left|u_1^2(t)-v_1^2(t)\right|
  +\frac{p}3\max\{u_1(t), v_1(t)\}~,
  \label{e.fidelity}
\end{align}
where $p=
|\mean{\exp(i\pi\sum_{n=2}^{N} c_n^{\dagger}c_n)}| = 
|\mean{\prod_{n=2}^{N} \left(-\sigma_n^z\right)}|$ is the expectation
value of the parity in the initial state $\ket{\Psi_{\tilde\Gamma}}$ 
of $\tilde\Gamma$.
When $\ket{\Psi_{\tilde\Gamma}}$ has a definite parity, $p=1$ and the
transmission quality only depends on the dynamical amplitudes $u_1(t)$ and
$v_1(t)$.
Several initial states that are experimentally achievable 
\cite{weitenberg2011single,Koetsier2008,Barmettler2008}
have a definite parity, 
notably the ground state, the fully polarized state, and the 
N\'eel state. On the other hand, when the chain $\tilde\Gamma$
is in a thermal state
with some inverse temperature $\beta$, then
$p=\prod_k \tanh(\beta\tilde E_k/2)$ where $\tilde E_k$ refers to the
energy eigenvalues $\tilde\Gamma$. For this reason, 
the transmission quality can be significantly suppressed 
above a certain temperature threshold. 
In the worst case scenario, when $p=0$, the fidelity is always
lower than the classical value \cite{horodecki1999general}, i.e. $F<2/3$,
and there is no benefit in using a quantum data-bus.
The non-local Wigner-string operator 
introduces entanglement between the boundary qubit
$N$ and the bulk which strongly suppresses the quality of transmission. 
In this regime the gap of the Hamiltonian sets the temperature threshold 
below which the initial state can be considered as the ground state, so
with $p\simeq 1$.  However, special attention is needed before
trying to use a gapped Hamiltonian for increasing the maximum temperature. 
Indeed, known formulas for the Hamiltonian gap are usually obtained in the 
thermodynamic limit and for closed boundary conditions.
It is important to stress that in open finite chains there could be
some differences:
for instance, the XY Hamiltonian is gapped for closed 
boundary conditions but there is an out-of-band mode in open chains 
which has exponentially small energy and makes the open XY Hamiltonian 
gapless \cite{LiebSM1961}. 
However, taking into account features specifically arising from open
boundary conditions, 
finite systems have generally a finite gap. Therefore,
given a sufficiently low temperature, 
the initial state can be considered a pure state with definite parity.
In sections~\ref{s.xx} and \ref{s.optimalxry}
this approximation is assumed, 
while the constraints on the initial state  are 
relaxed in Section~\ref{s.encoding}. 
%

\section{Ballistic quantum information transfer with XX models}\label{s.xx}
\subsection{Perfect state transmission}
In this section we study the state transmission in XX models, where
$\gamma_n=0$, the number of fermions in the initial states is conserved,
and $V(t) = 0$. We show how to make an XX spin chain a 
perfect mirror via a suitable engineering of the coupling strengths $j_n$. 
Then we introduce 
a minimal engineering scheme where all the couplings
are uniform $j_n=j=1$ ($j$ sets the energy scale) 
except at the boundaries, $j_1 = j_{N-1} \neq j$. 

Hamiltonians acting as perfect dynamical mirrors have been introduced 
in \cite{ChristandlDDEKL2005,albanese2004mirror}.
With {\it ad-hoc} engineering of {\it all } 
the interactions, one can perfectly transfer 
not only the states between the ends of the chain,
 but also whatever state
in position $x$ to the position $N-x+1$, $N$ being
the length of the chain. In order to understand what are the dynamical 
features of these chains let us 
consider an arbitrary initial state with $M<N$ spins in the state $\ket 1$ and 
$N-M$ spins in the state $\ket 0$. This state can be written as 
\def\mmmm{\{m_1<\cdots <m_M\}}
\def\mmml{\{\ell_1<\cdots <\ell_M\}}
\begin{equation}
  \ket\psi = 
  \sum_{\{m_1<m_2<\cdots<m_M\}}
  \psi(\{m_n\}) \sigma_{m_1}^+\sigma_{m_2}^+\cdots
  \ket 0 = \sum_{\mmmm}
  \psi(\{m_n\}) c_{m_1}^\dagger c_{m_2}^\dagger \cdots \ket 0~,
  \label{e.slater}
\end{equation}
where $\ket 0 = \ket{00\cdots}$ is the vacuum of the Fermi operators and where
the ordering $m_1 < m_2 < \cdots<m_M$ is required for
removing the action of the Wigner string. After a certain time $t$,
\begin{align}\nonumber
  \ket{\psi(t)} 
  &= \sum_{\mmmm}\sum_{\{\ell_n\}}
  \psi(\{m_n\}) U_{m_1 \ell_1}^*(t)U_{m_2 \ell_2}^*(t) ~
  c_{\ell_1}^\dagger c_{\ell_2}^\dagger \cdots \ket 0 \\&= 
  \sum_{\mmmm}\psi(\{m_n\}) \sum_{\mmml}
  \det\{U^*_{m_j,\ell_k}(t)\} 
  \sigma_{\ell_1}^+ \sigma_{\ell_2}^+ \cdots \ket 0~,
\end{align}
where $\det\{U^*_{m_j,\ell_k}(t)\} = 
\sum_\pi (-1)^\pi \prod_j U^*_{m_{\pi(j)} \ell_{j}}(t)$ and the
$\pi$'s refer to the permutation of the indices $\{\ell_n\}$.
In the last equality we have chosen a proper order of the indices for mapping
the many-fermion state back to the spin representation.
A perfect transmission occurs if at a certain time $t^*$ it is
\begin{align}
  U(t^*) = e^{i\alpha} X~,&& X_{nm}\equiv\delta_{n,N-m+1}~,
  \label{e.reflec}
\end{align}
where $X$ is the reflection operator and $\alpha$ an arbitrary 
phase. If this is the case $\ket{\psi(t^*)}$ becomes
\begin{align}
  \ket{\psi^{\rm mirrored}} = 
  \sum_{\mmmm}
  e^{i\alpha_{\{m_n\}}} \,\psi(\{N+1-m_n\})\,
  \sigma_{m_1}^+\sigma_{m_2}^+\cdots \ket 0~,
  \label{e.stmirrored}
\end{align}
with a phase $\alpha_{\{m_n\}} = 
\frac\pi2 \bar n(\bar n-1) - \bar n \alpha$, where $\bar n$ is the number 
of fermions in the state. The phase $\alpha_{\{m_n\}}$ can be made constant
in every sector with a definite parity by choosing, e.g., $\alpha=\pi/2$.
In the XX model this can be done for instance with a proper choice of the 
magnetic field. 
Nevertheless, when $\ket\psi$ is a
superposition of states with different parities,
non-trivial effects can occur and the dynamics
can generate an entangling gate \cite{yung2005perfect,gate}.

An XX spin chain generates a dynamical mirror operator 
if Eq.~\eqref{e.reflec} is satisfied. 
We now analyse what are the Hamiltonians
whose dynamics satisfy Eq.~\eqref{e.reflec} for some $t^*$. 
In the XX model, as $B=0$ in Eq.~\eqref{e.Hfermi}, $U(t) = e^{-it A}$; so
the first requirement for perfect transmission is that the 
hopping matrix $A$ and the 
reflection matrix $X$ have to be diagonal in the same basis, i.e. 
\begin{equation}
  [A,X] = 0~.
  \label{e.persym}
\end{equation}
The above condition requires a mirror-symmetric Hamiltonian. Indeed, 
the hopping matrix $A$ has to be symmetric with respect to the
``anti-diagonal'', a property called \emph{persymmetry} in the mathematical 
literature \cite{CantoniB1976}: $j_n = j_{N-n}$ and $h_n = h_{N-n+1}$.
Furthermore, calling $\omega_k$ the eigenvalues of $A$, 
Eq.~\eqref{e.reflec} forces the existence of 
a time $t^*$ such that $e^{-i \omega_k t^*}$ is proportional to the eigenvalues 
of $X$. If the energy eigenvalues $\omega_k$ are sorted in increasing 
order, then one can show \cite{yung2005perfect} that Eq.~\eqref{e.reflec}
requires the following condition 
\begin{equation}
  e^{-i \omega_k t^*} = e^{i\alpha} (\minus 1)^k~,
  \label{e.mirrorsym}
\end{equation}
where $(-1)^k$ are the eigenvalues of $X$.
When the matrix $A$ is persymmetric and the
condition \eqref{e.mirrorsym} holds, then perfect transmission 
is obtained between sites which are at the same distance from
the opposite boundaries: the XX chain acts as a perfect dynamical
mirror. Condition \eqref{e.mirrorsym} can be solved 
numerically using inverse eigenvalue techniques, i.e. algorithms giving
an Hamiltonian with the required spectrum 
\cite{Parlett1998,bruderer2012exploiting}.
Moreover,
several analytic solutions of \eqref{e.mirrorsym} have been found 
\cite{kay2009review}.  The simplest amongst these 
\cite{Albanese2004} 
requires no magnetic fields, $h_n=0$, and 
a full engineering of the couplings according to the law
$j_n \propto \sqrt{n(N-n)}$.

To the best of the author's knowledge, all perfect state transfer schemes 
are based on fully engineered chains.  These approaches are thus 
overwhelmingly complicated in 
an experimental perspective, as an experimentalist should be able to perform
a fine tuning of the interactions according to, {\it e.g.},
the law $j_n \propto \sqrt{n(N-n)}$.
Moreover, the dependence on $N$ of the coupling strengths avoids scalability:
when the transmission distances are varied, all the
nearest-neighbour interactions have to be changed as well.

In the next sections different models are introduced where the condition 
\eqref{e.mirrorsym}, though not exactly satisfied for each $k$, 
is accurately fulfilled
 by those modes which are relevant for the dynamics. Although condition
\eqref{e.mirrorsym} sets the fundamental requirement for achieving perfect state
transmission with fully engineered models, it is also one of the main 
building blocks 
for ballistic quantum state transmission with minimally engineered models.

\subsection{Minimal-engineered models and wave-like dynamics}\label{s.minimal}

Transferring quantum states from sender to receiver does not require the
full mirror inversion of the whole chain. 
What is really needed is the swap of the boundary states 
irrespective of what happens to the rest of the chain. Therefore, condition 
\eqref{e.reflec} is overblown. The only requirement is that
$u_1(t^*)=|U_{1N}(t^*)|\simeq 1$ for some transfer time $t^*$ which should be
as short as possible, i.e. 
$t^* = \mathcal O\left({N}/{\max_j |j_n|}\right)$. 
How can this goal be accomplished and the unnecessary hypotheses relaxed?
First of all it is known that the mirror-symmetry of the Hamiltonian,
i.e. Eq.\eqref{e.persym}, is still a necessary prerequisite for
the transmission between site $1$ to $N$ \cite{kay2009review}. 
We thus consider mirror-symmetric Hamiltonians and write
the spectral decomposition of the hopping matrix $A$ as
$A=O\omega O^T$. Then one can show 
\cite{banchi2011long} that
\begin{align}
  u_1(t) = \left|\sum_k O^2_{1k} (-1)^k e^{-i\omega_k t}\right|~,
  \label{e.wavez}
\end{align}
where the identity 
$O_{1k}=\pm(-1)^kO_{Nk}$, which is a property of persymmetric matrices 
\cite{CantoniB1976}, has been used.
The above equation shows that a high quality transmission can be obtained 
when the condition
\eqref{e.mirrorsym} is {\it approximately} satisfied by the modes $k$ which
mainly influence the dynamics, namely those for which 
$O^2_{1k}$ is significantly different from zero. 
This requirement can be fulfilled with a minimal engineering of the 
interactions. Minimally engineered models are a deviation from the
uniform case and the resulting  hopping matrix $A$ is quasi-uniform
\cite{banchi2013spectral}; therefore, 
the model can be diagonalized and its features expressed by
means of shifted quasi-momenta $q \approx \pi k /N + \eta_k$, where
$k=1,\dots,N$ and the shifts $\eta_k = \mathcal O(N^{-1})$ depend 
on the parameters of the non-uniform part of the Hamiltonian.
Using the quasi-momentum $q$ as an alternative index for $k$, one realizes 
that 
\begin{align}
  u_1(t) \approx \left|\sum_q \varrho(q) e^{i [N q - \omega_q t]}\right|~.
  \label{e.wave}
\end{align}
where $\varrho(q) = O_{1q}^2$. The above equation models a wave packet in 
momentum space which evolves with an energy $\omega_q$ that, being the model
quasi-uniform, can be interpreted as a dispersion relation. 
Although $\omega_q$ can be a complicated function, 
a ballistic coherent quantum information transmission
is obtained by designing
the wave-packet in momentum space in such a way 
that $\varrho(q)$ is peaked around the
inflection point of the dispersion relation, i.e. where 
$\omega\approx v q$ for some group velocity $v$. 
Indeed, in this case 
$u_1(t{\simeq}N/v) \approx \sum_q \varrho(q) = 1$.

A ``wave-packet encoding'' scheme
suitable for quantum communication has been proposed
also in \cite{OsborneL2004,Haselgrove2005}. However, in that case the
wave-packet is created in real space by encoding the state
to be transferred into a wave-packet over multiple sites of the chain.
Classical wave dynamics theory is then
exploited for designing the optimal shape
of the packet so as to minimize the dispersion.
On the other hand, the minimal-engineering  approach does not require the
control of multiple sites for the encoding, as the wave-packet is not
formed in the site-space but rather in the quasi-momentum space. 
Indeed, 
the non-uniform interactions make the system not
diagonalizable by a Fourier transform, and 
the resulting wave-packet shape $\varrho(q)$ can 
be tuned, as shown below, by acting on the non-uniform part 
of the Hamiltonian.

In \cite{Leonardo10,banchi2011long} the simplest Hamiltonian 
suitable for high-quality ballistic quantum state transmission 
has been found.  The model consists of an XX spin chain with no
applied magnetic field, $h_n = 0$, and 
\begin{align}
  j_n &= 1, \text{ for } n=2,\dots,N-2~, &
  j_1 &= j_{N-1} \neq 1~.
  \label{e.jjjjj}
\end{align}
This particular choice of minimal couplings 
is natural for applications. Consider for instance a potential 
implementation in a quantum device: 
sites $1$ and $N$ are part of two distant quantum registers, while the 
rest of the chain models the physical medium for connecting them. 
When the sender's state has to be transferred to the receiver,
the interaction $j_1$ is switched on and the (many-body) dynamics of 
the chain is then used for mediating the transmission.
As in a potential application the couplings $j_1=j_{N-1}$ must be controllable,
it is natural to require that they could also be switched on and set to a 
specifically tuned value which is different from the other couplings of
the chain.
The XX model with non-uniform interactions \eqref{e.jjjjj} can be 
solved analytically: the dispersion relation reads
\begin{equation}
 \omega_q = \cos{q} ~,
\label{e.omegak}
\end{equation}
where the quasi-momentum $q$ 
takes $N$ discrete values $q_k$ in the interval $(0,\pi)$: 
\begin{align}
 q_k &= \frac{\pi k + 2\varphi_{q_k}}{N{+}1}~, 
\label{e.x0.kn}
&
 \varphi_q &= q-\cot^{-1}\!\Big(\frac{\cot{q}}{\Delta}\Big)
 ~~~~\in\big({\textstyle -\frac{\pi}2,\frac{\pi}2}\big) ~,
& \Delta&=\frac{j_1^2}{2-j_1^2}~.
\end{align}
The corresponding shape of the
wave-packet $\varrho(q)$, namely the density of the excitations, is 
\begin{equation}
 \varrho(q)= \frac1{N{+}1{-}2\varphi_q'}
  ~\frac{\Delta(1{+}\Delta)}{\Delta^2+\cot^2{q}}~.
\label{e.x0.Parlett}
\end{equation}
Eqs.\eqref{e.omegak} and \eqref{e.x0.Parlett} show a peculiar property
of this quasi-uniform XX chain: the dispersion relation is almost linear around
the zero-energy zone and the density of the excitations $\varrho(q)$ is peaked
around this zone. The width of this peak is described by $\Delta$ and decreases
for decreasing $j_1$.

In the limit of very weak $j_1$ only the zero-energy modes 
are involved in the dynamics
\cite{wojcik2007multiuser,CamposVenutiBR2007,WojcikLKGGB2005,Yao2010}.
In this regime the dynamics is basically a resonant transmission mediated 
by one mode. Indeed, when $j_1\ll 1$ one can use perturbation theory
for tracing out the off-resonant modes and obtain an effective 
Hamiltonian acting on the boundaries 
\cite{wojcik2007multiuser}. The strength of the resulting long-distance 
effective
interaction is very weak (much weaker than $j_1$) and consequently 
the resulting transmission times are very long and non ballistic.

On the other hand, a coherent ballistic transmission must be a non-perturbative 
phenomenon. When $j_1$ increases, more and more
normal modes are involved and 
resulting dynamics emulates a wave packet, as in Eq.\eqref{e.wave}.
The fundamental observation is that a narrow 
$\varrho(q)$ is not the only requisite. Indeed,
as the model is not translationally invariant, 
the phase-shifts in the 
quasi-momenta can alter 
the dispersion relation 
for very weak $j_1$. The explicit calculation of the group velocity of the
wave-packet around the linear zone reads \cite{banchi2011long}
\begin{align}
  v \propto \partial_k\omega_{q_k} 
 = \frac{\pi}{t^*}\,\Big[1+
   \Big(2\,\frac{1{-}\Delta^2}{t^*\Delta^3}{-}\frac12\Big)
   \cos^2\!{q_k}+(\cos^4\!{q_k})\Big] \,,~~~~
   \label{e.enershift}
\end{align}
where $t^*\,{=}\,N{+}1+2\,(1{-}\Delta)/\Delta$ is the arrival time.
Being the wave-packet 
peaked around the linear zone, the first non-linearity comes from the
cubic terms of $\omega_q$ and depends on both $N$ and $j_1$.
The dispersive term can be minimized 
via an optimal choice of $j_1 = \mathcal O(N^{-1/6})$. This is the
reason for the existence of a \emph{non-weak} optimal value \cite{Leonardo10}
ensuring a coherent ballistic transmission.

From a quantitative point of view it is found that the optimal values
$j^{\rm opt}_1$ are not those minimizing the dispersive term 
in \eqref{e.enershift}. Indeed those optimal values come from a trade-off of
two competing requirements
\begin{enumerate}
  \item $j^{\rm opt}_1 < j^{\rm w}_1$, where $j^{\rm w}_1$ 
    represent the optimal 
    coupling for making the width of the wave packet significantly
    different from zero only around the linear zone.
  \item $j^{\rm opt}_1 > j^{\rm \ell}_1$ where $j^{\rm \ell}_1$ 
    represents the threshold 
    below which the ``linear zone'' is no more linear, as shown in 
    Eq.~\eqref{e.enershift}.
\end{enumerate}
These two competing effects cannot be efficiently 
managed via a single parameter $j_1$: 
the optimal coupling is a compromise, and is found 
to scale as $j_1^{\rm opt} = \mathcal O(N^{-1/6})$. When the optimal coupling
is used the obtained transmission fidelity is above $95\%$ for 
$N\approx 100$ and $F > 90\%$ even in the thermodynamic limit
\cite{Leonardo10,banchi2011long}.
The advantages of the coherent ballistic dynamics are evident: 
only the couplings $j_1$ need to be controlled, 
there is no need for 
engineering, nor for controlling many qubits for encoding a wave-packet and, 
thanks
to the non-weak couplings, transmission times are fast.

%

The two competing constraints discussed above are a general feature of 
minimally-engineered models and in \cite{apollaro201299} it has been shown
that they can be overcome by introducing another optimally tuned value
$j_2 = j_{N-2}$. We will discuss this point in detail in 
Section~\ref{s.encoding}, while 
in the next section we study how to induce a ballistic quantum information
transfer in particle non-conserving models. 
%

\section{Minimal engineering of the XY model} \label{s.optimalxry}
We have shown that the problem of transferring a quantum state in an XX
spin chain is equivalent to the transmission of a fermion in one 
dimension and we have optimized such transmission using minimal requirements.
The main guideline of the optimization procedure 
is to tune the interactions such
that the travelling fermion generates a wave-packet whose shape in the
quasi-momentum space
can be controlled by the boundary couplings. By tuning that 
shape around the linear zone of the dispersion relation a coherent ballistic 
dynamics occurs. In this section we extend this approach to the XY spin model, 
whose Hamiltonian can be mapped into the fermionic Hamiltonian 
\eqref{e.Hfermi}. Unlike the XX case, the total magnetization along the $z$ 
direction is not a constant of motion and, as $B\neq 0$, 
the dynamics does not conserve 
the number of fermions in the initial state. 
Nonetheless the model can still be diagonalized by a canonical Bogoliubov
transformation, and written into a set of independent (non-interacting) 
fermionic modes. With some technical modifications the minimal optimization 
procedure can be applied even in this case.

As in the previous section, before introducing the simplest engineering, 
we state some general comments about the properties
that an XY Hamiltonian should satisfy in order to act as a (quasi) perfect 
quantum state transmitter.
For the sake of simplicity, we consider $\gamma_n=\gamma\,j_n$.
An XY spin chain acts as a perfect mirror if 
$U(t^*) = e^{i\alpha} X$, for some $\alpha$ and $V(t^*) = 0$ at the 
transmission time $t^*$; 
models where this condition occurs have been
studied in the literature \cite{di2008perfect,di2008control},
and are known to require 
the full engineering of the couplings $j_n$, plus 
a further control on the local magnetic field $h_n$.

Proceeding as in the XX case we here study how 
the conditions required for a perfect mirroring of the whole chain 
can be reduced if one rather aims at obtaining a high quality state
transmission.
In Appendix~\ref{s.quaddiag}, and in particular in Appendix~\ref{s.xycondition},
we analyse the role of the symmetries. 
It is shown that the mirror-symmetry is a fundamental condition both for
operating as a perfect dynamical mirror and for transferring 
states between sender and receiver.
Therefore, we assume the mirror-symmetry also for the XY spin chain.
Due to this symmetry, as shown in Appendix~\ref{s.fermiJ}, it is 
\begin{subequations}
  \begin{align}
    U_{1N}(t) &= \sum_k \left[ \left(\frac{W_{k1}+s_kW_{kN}}2\right)^2 s_k\,
      e^{-i\omega_k t}
      - \left(\frac{W_{k1}-s_kW_{kN}}2\right)^2 s_k e^{i\omega_k t}\right]~, \\
      V_{1N}(t) &= \sum_k \left(\frac{W_{kN}^2-W_{k1}^2}4\right) s_k
      \left(e^{-i\omega_k t}-e^{-i\omega_k t}\right)~,
    \end{align}
    \label{e.UVxy}
\end{subequations}
where $W$ is an orthogonal matrix required to define the canonical
transformation that diagonalizes the Hamiltonian,
$\omega_k$ are the corresponding energies, and $s_k = (-1)^k$. 
It turns out
that $W_{k1}\simeq \pm (-1)^k W_{kN}$. These relations are exact in the 
XX case and are true, though with some approximation,
also in the XY case in the most relevant cases, namely when it is found
$U_{N1}(t^*)\approx 1$ and $V_{N1}(t^*)\approx 0$
at the transmission time $t^*$. Consequently, the approximated
evolution operator $u_1(t) = |U_{N1}(t)|$ is 
\begin{align}
  u_1(t) \approx \left|\sum_k \varrho(k) e^{i(\pi k - \omega_k t)}\right|~, &&
  \varrho(k) = W_{k1}^2~.
  \label{e.approxu}
\end{align}
The above equation proves that, due to the mirror symmetry, 
the dynamics can be described as a wave-like evolution
travelling from sender to receiver also in the XY case. Although 
approximated, Eq.~\eqref{e.approxu}
has the same form of Eq.~\eqref{e.wavez}, which is valid for
the XX case. The difference stems from the shape 
of the wave-packet in the momentum space $\varrho(k) = W_{1k}^2$ which 
is slightly more complicated because of the different diagonalization 
procedure (see Appendix \ref{s.fermiJ}). 
In the following we investigate how one can
obtain an almost perfect transmission in XY chains.
Some numerical findings about XY spin chains that operate as coherent ballistic
channels have already been obtained in \cite{banchi2011efficient,Leonardo10}.


\begin{figure}[t]
  \centering
\resizebox{0.55\textwidth}{!}{%
  \includegraphics{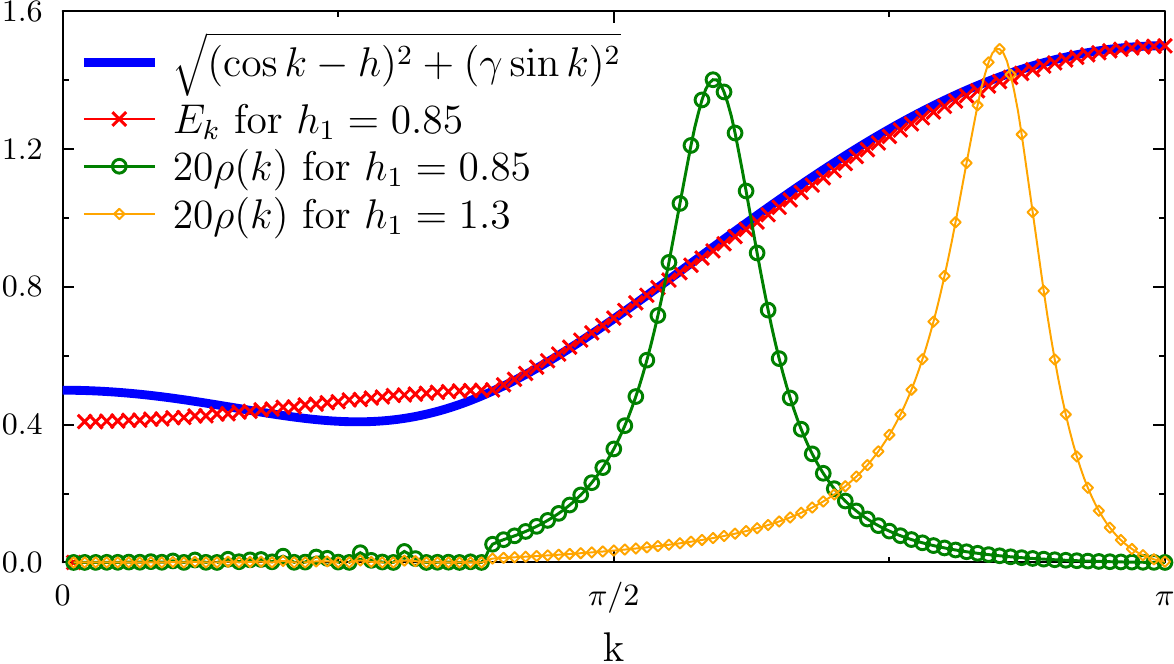}}
  \caption{
  Dispersion relation, energies $\omega_k$ and density of excitations for $N=100$,
  $\gamma=h=0.5$ and $j_1=j_{N-1} = 0.48$. The peak of the density $\varrho(k)$ 
  can be controlled by the local magnetic field: it turns out that the energy
  value of the peak is basically given by $h_1 = h_N$, i.e. by the local 
  energy. The seemingly bad agreement between $\omega_k$ and the 
  dispersion relation, in the left side, 
  is due to the numerical increasing ordering of the eigen-energies $\omega_k$.
  }
  \label{f.omegak}
\end{figure}

The achievement 
of a coherent dynamics in the XY model begins, as in the XX case,
with the analysis of the dispersion relation in the infinite chain limit,
\begin{equation}
  \omega_k\,{=}\sqrt{(h{-}\cos{k})^2+\gamma^2\sin^2k}~, 
  \label{e.disprelxy}
\end{equation}
displayed
in Fig.~\ref{f.omegak}. One can easily spot the existence of a 
region of linearity in the neighborhood of the inflection
point(s) $k_0$, where $\omega_k\approx \omega_{k_0}+v(k-k_0)$. 
It turns out that a local magnetic field 
$h_1\,{\simeq}\,\omega_{k_0}$ allows the peak of $\varrho(k)$ to be
centered around $k_0$ (see e.g. Fig.~\ref{f.omegak}). 
The physical reason behind the possibility of changing the peak position 
via $h_1$ is more clear in the weak coupling limit. 
Indeed, when $j_1\ll 1$ only the 
modes which are almost resonant with the local energy $h_1$ of the 
external qubits are involved in the dynamics. On the other hand,
when $j_1$ increases more modes come into play and 
the peak is found to widen without changing its position.
An optimal width can be determined numerically by finding the optimal
coupling $j_1^{\rm{opt}}$.
An optimal coupling is indeed expected to emerge, as in 
the XX case, from a compromise
between making $\varrho(k)$ peaked around $k_0$ and minimising the 
non-linearities introduced in $\omega_k$ around $k_0$ by the non-uniform 
interactions. 

There are two main differences between XX and XY models. The first one is
that, in the latter case, 
the region of linear dispersion depends on the parameters ($\gamma$ and $h$):
it sensibly shrinks as the anisotropy $\gamma$
increases (which might make the bus useless) and 
can be extended by increasing the field $h$.
For example, in the ``extreme'' case of
the Ising chain ($\gamma\,{=}\,1$), when $h\,{=}\,0$ the dispersion
relation becomes flat and does not allow for propagation. 
This explains the observation~\cite{Bayat2008a} that in
such limit no entanglement propagation takes place: indeed, a
vanishing group velocity means that nothing can be transmitted over
the chain.  However, a
wide linear region can yet be obtained by applying a finite magnetic field
$h$ and one can act on the latter parameter so as to fulfil the
conditions for optimal dynamics. 
The second important difference is the need for an extra parameter, i.e. 
the local magnetic field. 
In the XX case (say for $h=0$) we have seen that 
the inflection point $k_0\,{=}\,\pi/2$
corresponds to $\omega_{k_0}\,{=}\,0$. There is no need for an extra local
magnetic field because $\varrho(k)$ is already peaked around the mode with
zero energy, which is notably also the center of the linear zone. 
On the other hand, the dispersion relation of the XY model \eqref{e.disprelxy} 
is gapped and 
$\omega_{k_0}\neq 0$. Unlike the XX case, when the XY model is considered, 
one has to
switch on a local magnetic field $h_1\,{\simeq}\,\omega_{k_0}$ in
order to increase the average energy of the initial state and make
$\varrho(k)$ peaked around the linear zone.
In XY models the analytical expressions are more complicated as the 
Hamiltonian cannot be expressed in terms of symmetric 
quasi-uniform tridiagonal matrices 
\cite{banchi2013spectral}, so the optimal parameters have to be 
found numerically \cite{banchi2011efficient,Leonardo10}. In the following
we analyse the transmission in an Ising model ($\gamma=1$) 
where, on the other hand, some analytical calculations can be performed.

\subsection{Ising case}
The Ising chain with non-uniform interactions can be analytically
diagonalized: the matrix $Z \, Z^T$, being $Z=A-B$ 
(see Eq.~\eqref{e.Hfermi} and appendix \ref{s.quaddiag}), 
is a quasi-uniform tridiagonal matrix with non-uniform corners. 
Thanks to the results of \cite{banchi2013spectral} one can show that
for $N\gg1$ the eigenvalues of $Z^T\,Z$ are 
$\omega_k^2 = 1+h^2-2h\cos k$, i.e. the square of the dispersion
relation \eqref{e.disprelxy}. Taking the quasi-momentum operator $k$ as
an alternative index one can show that 
\begin{align}
  \varrho(k) &\propto
  ~\frac{j_1^2\sin^2\!{k}}
 {[(2{-}j_1^2)\cos{k}-x]^2+j_1^4\sin^2\!{k}}~,
 &
  x &=  \frac{1+h^2-h_1^2 - j_1^2}h.
\label{e.rhokising}
\end{align}
The peak of $\varrho(k)$ is obtained for 
$k_0=\cos^{-1}\frac{x}{2-j_1^2}$, while
the energy of the mode $k_0$ is 
\begin{equation}
  \omega_{k_0} = 
  \sqrt{1+h^2-2h\cos k_0} = \sqrt{\frac{2h_1^2+(1-h^2)j_1^2}{2-j_1^2}} ~,
  \label{e.erisonising}
\end{equation}
and 
the width of the peak is given by 
\begin{equation}
  \Delta=\frac{j_1^2}{\sqrt{(2 - j_1^2)^2 - x^2}}~.
\end{equation}
In the limit $j_1\rightarrow 0$ the resonant mode is the one with
energy $|h_1|$, as expected; however, for finite $j_1$, 
a shift is observed.
Moreover, it turns out that 
  $\left|W_{kN}\right| = \frac{h_1}{\omega_k}\left|W_{k1}\right|$; 
when $h_1 \simeq \omega_{k_0}$, as we are interested in the neighbourhood
of the resonant mode, we can neglect the prefactor and consider 
$\left|W_{k1}\right| \simeq \left|W_{kN}\right|$: this is an analytical
proof of the validity of the approximated formula \eqref{e.approxu}.

\begin{figure}[t]
 \centering
 \resizebox{!}{0.18\textheight}{%
    \includegraphics{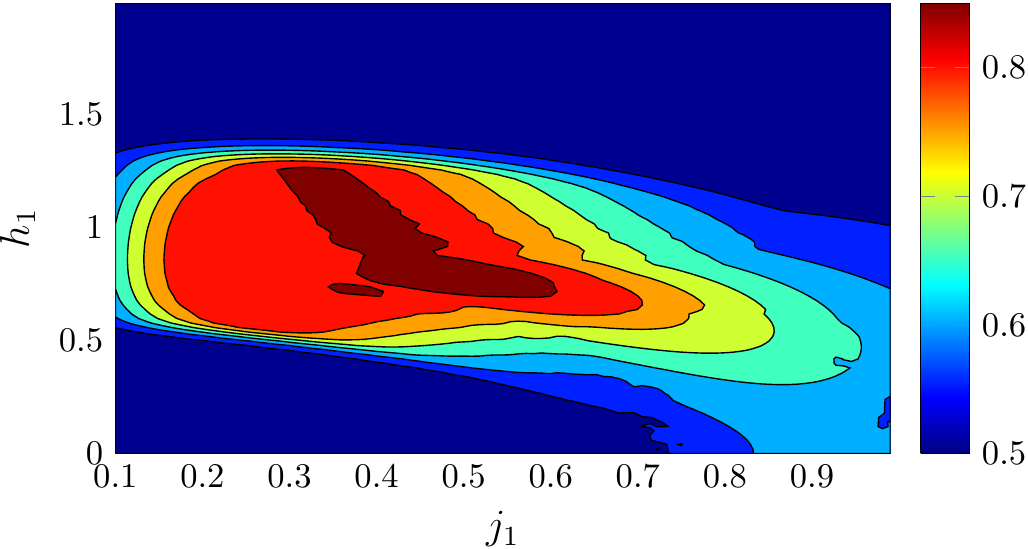}}
    \resizebox{!}{0.18\textheight}{%
  \includegraphics{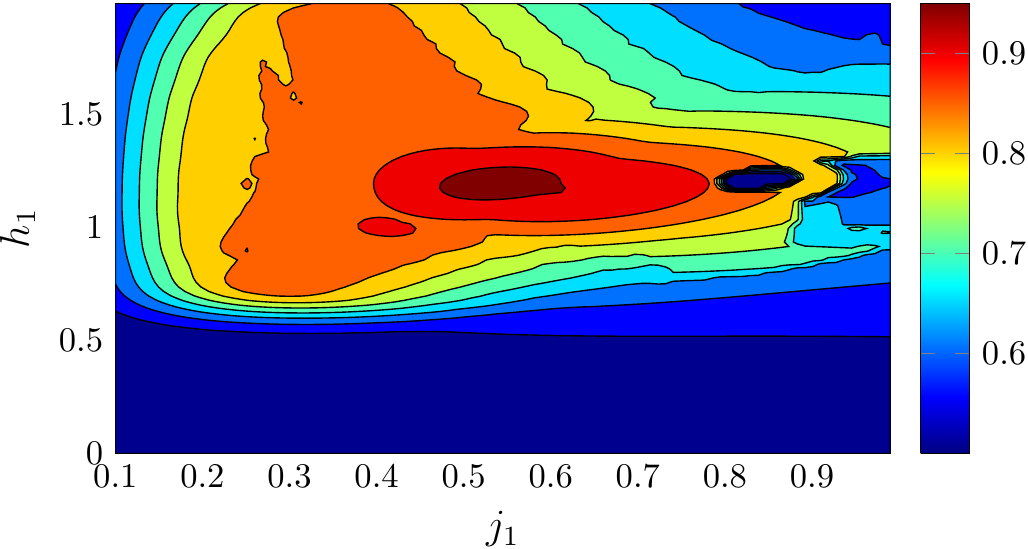}}
  \caption{ Contour plot of the fidelity of transmission $F(t^*)$ for $N=50$,
    $\gamma=1$, $h=0.5$ (left) and $h=1.5$ (right). The transmission times
    have been obtained by maximizing the $F(t)$ for ballistic times, i.e.
    when $t\approx N/v$, being $v=\min\{h,h^{-1}\}$ the group velocity
    around the linear zone. The plots show the optimal value of $j_1$ and 
    $h_1$ and the corresponding high fidelity.
  }
 \label{f.jh}
\end{figure}
\begin{figure}[t]
 \centering
 \resizebox{0.5\textwidth}{!}{%
    \includegraphics{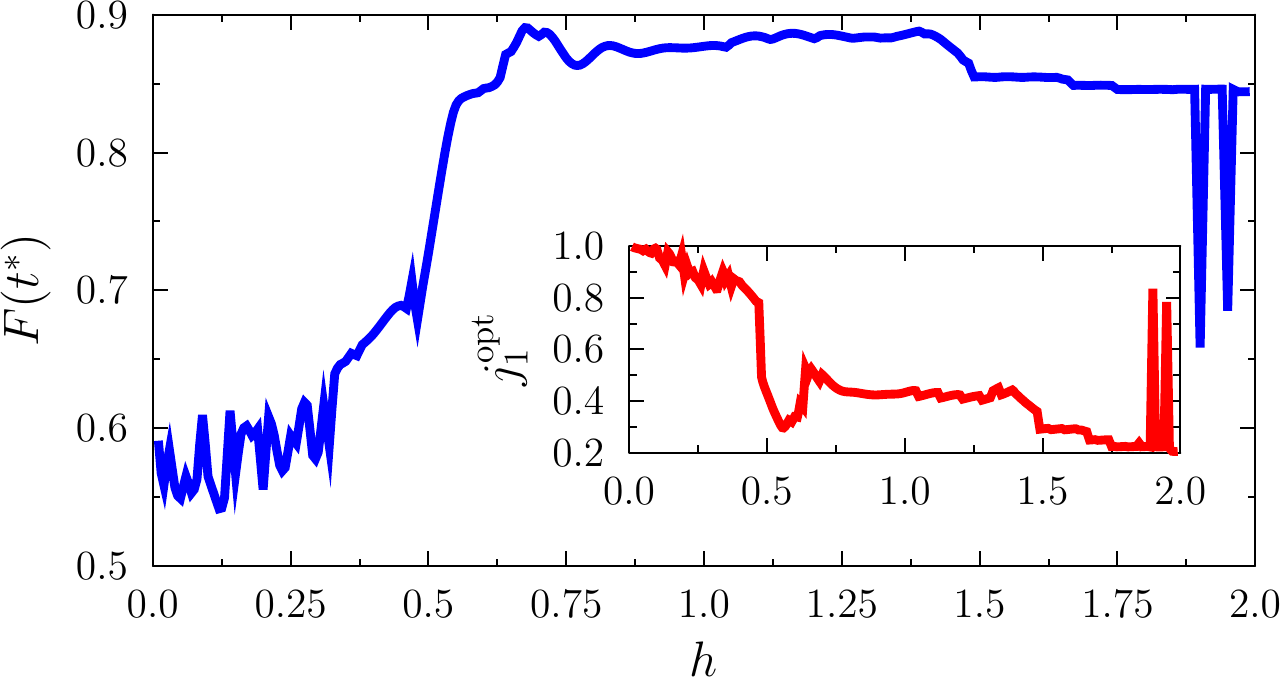}}
  \caption{ Transmission fidelity for different $h$, $N=50$, $\gamma=1$ and
    $h_1=h$. The inset shows the corresponding optimal values $j_1^{\rm opt}$.
  }
 \label{f.ising}
\end{figure}

For making the optimal dynamics to emerge, the position of the peak must be 
centered around the linear zone of the dispersion relation 
$\bar\omega=\sqrt{|1-h^2|}$, i.e. $\omega_{k_0}=\bar\omega$. 
This condition is satisfied by setting
\begin{equation}
  h_1 = \begin{cases}
    \sqrt{(1 - j_1^2) (1 - h^2)}  & \text{~ ~ ~ for ~ ~ ~} |h|\le 1~,\\
    \sqrt{h^2-1}  & \text{~ ~ ~ for ~ ~ ~} |h|\ge 1~,\\
  \end{cases}
  \label{e.h0ising}
\end{equation}
and the corresponding width is
\begin{equation}
  \Delta = 
  \frac{\max\{|h|,1\}}{\sqrt{|1-h^2|}}\;\frac{j_1^2}{2 - j_1^2}~.
  \label{e.widthising}
\end{equation}
The above reasoning cannot be applied at the ``critical'' value $h=1$, as it
predicts $h_1=0$. However, $h_1=0$ corresponds to $W_{kN}=0$,
$u_1(t) = v_1(t)$ and yields a very low transmission fidelity.
Notice that the width $\Delta$ has a 
prefactor, as compared with the XX case, which is greater than 1. 
The optimal $j_1$ is then expected to be smaller than the XX counterpart.

In Fig.~\ref{f.jh} we plot $F(t^*)$ for different parameters and
show that the fidelity for the optimal values of $j_1$ and $h_1$ exceeds
$95\%$ for a chain of 50 spins when $h=1.5$ while it is slightly lower for
$h = 0.5$. In both cases the optimal values are quite in agreement with
the estimate \eqref{e.h0ising}, although the latter has been
obtained after different approximations.

The above analysis has been obtained by looking only at the density of
the excitations, without considering the perturbations of the dispersion
relation due to the local non-homogeneous coupling $j_1$ and magnetic field
$h_1$. However, these non-homogeneous terms introduce a shift in the 
quasi-momenta \cite{banchi2013spectral} that alters the dispersion relation.
In the XX case, we have seen that this effect can 
transform the linear zone into a dispersive one.
In principle, 
even the opposite could happen: the quasi-momenta shift due to the 
non-homogeneous interactions might 
linearize $\omega_k$ around a certain $k_0$, ultimately
making the dynamics more coherent.
We have numerically observed that the phase shift mostly affects 
$\omega_k$ around the peak $k_0$ of $\varrho(k)$. Therefore, we have performed
a numerical simulation without considering the optimization of $h_1$: even
if the resulting $\varrho(k)$ is not peaked around the linear zone, the
subsequent optimization of $j_1$ can slightly linearize $\omega_k$ around
$k_0$.
The results of the numerical analysis with $h_1=h$ are shown in 
Fig.~\eqref{f.ising}.
As expected, the transmission fidelity is lower compared to the case where
even the position of $k_0$ is optimized via $h_1$.
 Nonetheless, for $h>0.6$ the transfer quality is considerably high.

\section{Minimal encoding for reliable transmission regardless of 
temperature and initial state}
\label{s.encoding}

The minimal engineering scheme introduced in the previous section has the 
advantage of requiring only the ability to address the boundary qubits and
to switch on the couplings between the boundaries and the bulk to an optimal 
non-perturbative value. With these minimal requirements an XY spin chain acts
as a reliable transmission channel with fast transmission times
$t^*=\mathcal O(N/j)$. Depending on the physical implementation,
the optimal dynamics can be improved if one can address and manipulate 
the couplings between two further
qubits, i.e. if one can operate locally on the qubits $1,2,N-1,N$. 
In \cite{apollaro201299} indeed it has been shown that 
the two constraints for obtaining the optimal dynamics
discussed in Section~\ref{s.minimal} can be better satisfied 
by tuning a second parameter $j_2=j_{N-2}$.
Provided that 
the initial state of the chain $\tilde\Gamma$ has a definite parity,
by properly tuning $j_1=\mathcal O(N^{-1/3})$ and  
$j_2=\mathcal O(N^{-1/6})$ one
obtains a fidelity higher than $99\%$ for $N\to\infty$.
This two-coupling engineering is not just a step towards the full 
engineering of the chain: the need for a further parameter 
arises from the requirement that two 
constraints have to be satisfied. 

The ability to address four qubits (two on the left and two on the right) 
also permits one 
to disentangle the transmission from the initial state 
of the internal chain \cite{MarkiewiczW2009} and then
make the transmission independent of $\ket{\Psi_{\tilde\Gamma}(0)}$.
Although XY models are quasi-free, the spin operator 
$\sigma^{-}_N = \Pi\,c_N$ is non-local in the fermionic picture because
of the parity operator $\Pi$. 
The latter makes the transmission strongly dependent on the initial state
of the bulk \cite{banchi2011long}: if $\ket{\Psi_{\tilde\Gamma}(0)}$ 
does not have a definite
parity, then there is a destructive interference which strongly suppresses the 
transmission. 
With a suitable four-qubit protocol, such a
destructive interference 
is cancelled. 
Similar schemes have also been 
proposed for achieving quantum state transfer in systems at
high (infinite) temperature
\cite{Yao2010,yao2013quantum,cappellaro2011coherent,ajoy2012mixed}.

We briefly discuss here the protocol introduced in \cite{MarkiewiczW2009}, 
while in the 
following section we extend this algorithm to make it suitable for
ballistic state transfer schemes.
Consider the initial state
$\rho_1(0) = (1+\vec r_1\cdot\vec \sigma)/2$,
parametrized by the Bloch vector $\vec r_1$.
This state can be encoded onto the state $\rho_{12}(0)$ of qubits 1 and 2
in two different ways $(\pm)$
\begin{align}
  \rho^\pm_{12}(0) &= \frac{I_{\pm,12} + r_1^x \sigma^x_{\pm,12} + 
  r_1^y \sigma^y_{\pm,12}+ r_1^z\sigma^z_{\pm,12}}2~,
  \label{e.encod}
\end{align}
\begin{align}
  \rho^-_{12}(0) &= \frac12\begin{pmatrix}
    0&0&0&0\\
    0&1+r_1^z&r_1^x-ir_1^y& 0\\
    0 & r_1^x+ir_1^y&1-r_1^z&0\\
    0&0&0&0
  \end{pmatrix}~,
  &
  \rho^+_{12}(0) &= \frac12\begin{pmatrix}
    1+r_1^z&0&0&r_1^x-ir_1^y\\
    0&0&0&0\\
    0&0&0&0\\
    r_1^x+ir_1^y&0&0&1-r_1^z\\
  \end{pmatrix}
  \label{e.initencoding}~,
\end{align}
where the above matrices are written in the computational basis 
$\ket{\alpha\beta}$, $\alpha,\beta\in\{0,1\}$, and 
where $\pm$ concerns the parity of the two-qubit state: $\rho^+_{12}$ 
(respectively $\rho^-_{12}$)
refers to the state encoded into the subspace where
$\sigma^z_1\sigma_2^z$ is positive (negative). 
The operators encoding the qubits according to the above formulae read
\begin{align}
  I_{\pm,12} &= \frac{1\pm\sigma_1^z\sigma_2^z}2~,&
  \sigma^x_{\pm,12} &= \frac{\sigma^x_1\sigma^x_2\mp\sigma_1^y\sigma_2^y}2~,\\
  \sigma^y_{\pm,12} &= \frac{\sigma^y_1\sigma^x_2\pm\sigma_1^x\sigma_2^y}2~,&
    \sigma^z_{\pm,12} &= \frac{\sigma^z_1\pm\sigma_2^z}2~.
\end{align}
When perfect state-transfer Hamiltonians are used \cite{MarkiewiczW2009} 
the state $\rho^\pm_{12}$ 
is perfectly transferred to the opposite end; at the transmission time
$t^*$ one obtains $\rho_{N,N-1}(t^*) = \rho^\pm_{12}$. The encoded
state \eqref{e.encod} is local in the fermionic picture and the transmitted
state depends only on the operators $c_N(t^*)$ and $c_{N-1}(t^*)$, 
as well as on their
Hermitian conjugate; having mapped the evolution into the transmission of two
fermions the parity is automatically removed.
Nevertheless,
when the quality of transmission is not perfect, the transmitted state
$\rho_{N,N-1}(t^*)$
may not have a definite two-qubit parity, making the decoding procedure 
more intricate. In the following part 
we discuss how to evaluate the transmission fidelity for this protocol.

\subsection{Fidelity of transmission for two-qubit encoding state transfer}
\label{s.2fid}
Before discussing the figure of merit for the transmission quality 
let us comment on the operations
needed for effectively realising the two-qubit encoding. If the initial state
of qubit 2 is $\ket 0$ (respectively $\ket 1$) then the encoding of the
initial state into $\rho^-_{12}$ (respectively $\rho_{12}^+$) is obtained 
simply using the CNOT gate: ${\rm CNOT} = \frac{1+\sigma^z}2\otimes 1+
\frac{1-\sigma^z}2\otimes\sigma^x$. In order to better illustrate the 
decoding procedure,
without loss of generality
we assume that the initial state of qubit 2 is the completely mixed state;
different initial states can be tackled in a similar way. Let $\mathcal C^\pm$ 
be the quantum operation realising the encoding, i.e.
\begin{align}
  \rho^\pm_{12} &= \mathcal C^\pm\left[\rho_1(0)\otimes\frac{1_2}2\right] ~,
  &\Longrightarrow&& \sigma_{\pm,12}^\alpha &= \mathcal C^\pm\big[
    \sigma_1^\alpha\otimes1_2\big]/2~.
  \label{e.coding}
\end{align}
Such a quantum operation cannot be a unitary gate, as the eigenvalues of the
input and output operators are different (two eigenvalues of $\rho^\pm_{12}$ 
are null). Let us consider for instance the preparation of $\rho^+_{12}$:
One can measure $\sigma^z_2$, flip the state of qubit 
$2$ depending on the result,
and then apply the unitary CNOT gate. Another possibility, which does not 
involve measurements, is to apply a local completely positive 
map $\mathcal P$ such that
$\mathcal P\big[\rho_1\otimes\rho_2\big] = \rho_1\otimes\ket1\bra1$ and then
apply a CNOT gate. In this particular case the encoding reads
\begin{align}
  \mathcal C^+\big[\rho\big] = 
  \sum_{i=1}^2 {\rm CNOT} \left(1\otimes A_i\right)\,\rho\,
  \left(1\otimes A_i^\dagger \right)  {\rm CNOT}  \label{e.C}
\end{align}
where we have chosen a particular set of Kraus operators realising the 
map $\mathcal P$: $A_1 = \frac{1+\sigma^z}2$, and $A_2 = \sigma^-$.

\begin{figure}[t]
  \centering
  \resizebox{0.55\textwidth}{!}{%
    \includegraphics{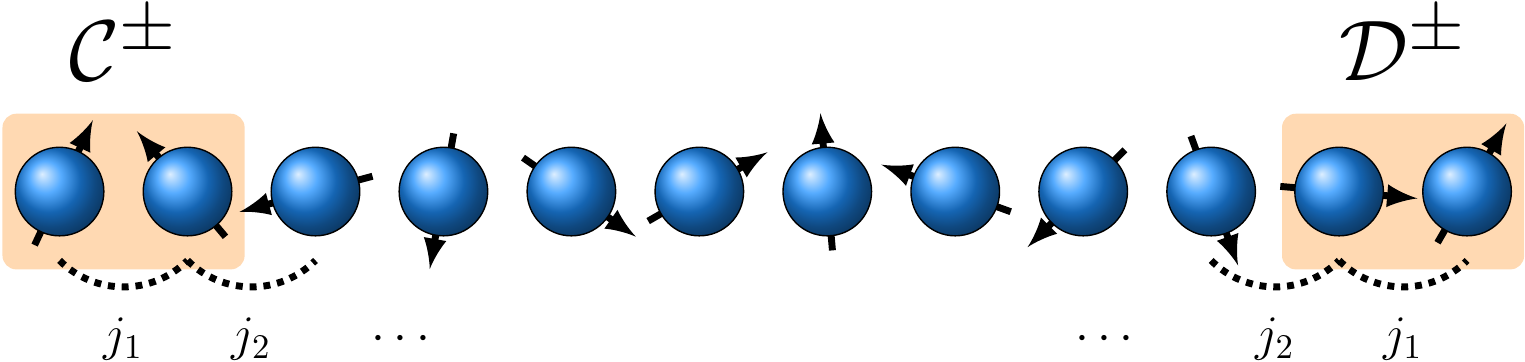}}
    \caption{Schematic representation of the state transfer protocol with the
      two-qubit encoding. 
      With the proper coding operation $\mathcal C^\pm$ and decoding operation 
      $\mathcal D^\pm$ one can make the result of the transmission
      independent of the initial state
      of the spins $2,\cdots,N-1$. The couplings $j_n=j$ are uniform except that
      at the boundaries where they take the value $j_1$ and $j_2$.
    } 
    \label{f.CD}
\end{figure}

We now investigate the necessary decoding quantum operation.
As shown in Section~\ref{s.quasifree}, 
in order to evaluate the transmission fidelity one has to obtain 
the matrix elements $D_{\alpha\beta}(t)$ which define the linear 
relation between the initial Bloch vector, with elements
$r_1^\beta(0)$, 
and the final evolved Bloch vector, whose elements
$r_N^{\alpha}(t) = \langle \sigma^\alpha_N(t)\rangle$
describe the state of the last qubit at time $t$.
After having applied the encoding \eqref{e.C} we suppose that the elements
$D_{\alpha\beta}(t)$ can be made independent of the parity, and in general
independent of $\ket{\Psi_{\tilde\Gamma}(0)}$, via a suitable 
decoding quantum operation $\mathcal D^\pm$. Let us write 
$r_N^\alpha(t)$ explicitly:
\begin{align}
  r_N^\alpha(t) = \Tr[\sigma_N^\alpha \rho(t)] = 
  \Tr\left(\sigma_N^\alpha\mathcal D^\pm_{N,N-1}\big[e^{-iHt} 
    \mathcal C^\pm_{12}\big[\rho(0)]e^{iHt}\big]\right)
   \equiv 
   \Tr\left({e^{iHt}
   \tilde {\mathcal D}}_{N,N-1}^\pm\big[\sigma_N^\alpha\otimes 1\big] e^{-iHt} 
   ~ \mathcal C^\pm_{12}\big[\rho(0)]\right)~,
  \label{e.rprimo}
\end{align}
where $\tilde{\mathcal D}$ is the dual quantum operation, i.e. if 
$\mathcal D[\rho]= \sum_i D_i\rho D_i^\dagger$ then 
$\tilde{\mathcal D}[\rho]= \sum_i D_i^\dagger \rho D_i$. The subscripts in
the above equation are used when we want to make explicit on which qubit the
codes act; moreover we frequently switch from the Heisenberg to the 
Schr\"odinger picture.
In principle it would be tempting to set 
$\tilde{\mathcal D}^\pm = \mathcal C^\pm$. Indeed, if this were possible
then $ {\tilde {\mathcal D}}_{N,N-1}^\pm\big[\sigma_N^\alpha\otimes 1\big]
= \sigma^\alpha_{\pm,N,N-1}$, namely one could write 
$r_N^\alpha(t) = \langle\sigma^\alpha_{\pm,N,N-1}(t)\rangle$ and make the
map independent of the parity of the chain. However, this is not feasible 
in general: For instance if one sets $\tilde{\mathcal D}^+ = \mathcal C^+$
with the encoding  \eqref{e.C}, then the physical operation 
$\mathcal D^+$, which is the real operation performed onto the state,
would not be a physical trace-preserving completely positive map. 
One could implement the decoding by measuring the qubit $N-1$, 
in a way similar to what has been described for realising 
the encoding operation. However, in this case
the mapping is not linear, as in  Eq.~ \eqref{e.affine}, 
and it may be much more complicated to 
obtain the transmission fidelity.

We consider here a simple decoding procedure. 
We do not claim that this is the optimal procedure, and we argue that, for 
increasing the fidelity, one could
find a better combination of local encoding/decoding protocols such that 
the encoded and decoded two-qubit states depend only on pairs of 
local fermionic operators and not on the parity.
Here we assume that the initial state is encoded into the state
$\rho_{12}^+$. As for the decoding operation, we simply apply
the CNOT gate to the qubits $N$ and $N-1$. With a long but straightforward
calculation one can evaluate the matrix $D(t)$ which describes the transmission
channel (see Section~\ref{s.quasifree})
\begin{align}
  D(t) = \begin{pmatrix} 
    \Re[U_{N,1}(t) U_{N-1,2}(t) -U_{N-1,2}^2] & 
    \Im[U_{N,1}(t) U_{N-1,2}(t) -U_{N-1,2}^2] & 
    \Re[(U_{N,1}(t) + U_{N-1,2}(t))U^*_{N-1,2}] \\
    -\Im[U_{N,1}(t) U_{N-1,2}(t) -U_{N-1,2}^2] & 
    \Re[U_{N,1}(t) U_{N-1,2}(t) -U_{N-1,2}^2] & 
    \Im[(-U_{N,1}(t) + U_{N-1,2}(t))U^*_{N-1,2}] \\
    0&0& |U_{N,1}(t)|^2+|U_{N-1,1}|^2 
  \end{pmatrix}~,
  \label{e.D}
\end{align}
where we have used the property $U_{1,N-1} = U_{2,N}$ which follows from 
mirror symmetry. Note that $\det D(t) > 0$ for every time $t$. On the other
hand, mappings such that $\det D(t) < 0$ are not suitable for transferring
quantum information, as the resulting fidelity is always lower than the 
classical value \cite{badziag2000local}. The final analytical expression for
the optimal transmission fidelity $F_{\rm e}$ for the two qubit protocol
is rather complicated. An approximated 
formula can be obtained assuming the optimal dynamics: as the aim of the 
optimal dynamics is to maximise $|U_{1,N}(t^*)|$ one can assume that 
$|U_{1,N-1}(t^*)|\approx 0$. Thanks to this approximation one obtains
\begin{align}
  F_{\rm e}(t^*) &\simeq \frac12+u_{1}(t^*)\left(
  \frac{u_{1}(t^*) + 2u_{2}(t^*)}6\right)~, &
  u_1(t) = |U_{1,N}(t)|~,\quad\quad
  u_2(t) = |U_{2,N-1}(t)|~.
  \label{e.Fapprox}
\end{align}
The above expression shows that the maximization of $u_1(t^*)$ is, as expected,
not enough for the two-qubit encoding protocol: since two states have to be
transferred to the opposite side, the state of qubit 1 to qubit $N$ and the 
state of qubit 2 to qubit N-1, one also has to find a proper scheme for 
maximizing $u_2(t^*)$ at the same time. In the following section we will
use a minimally engineered chain for achieving this goal.

\subsection{Minimal engineering for two-qubit state transfer}
\label{s.minimal2}

We have shown that, thanks to the two-qubit encoding, an XX spin chain 
can be used as reliable communication channel 
irrespective of the initial state of the chain, provided that
it is able to transfer the state of qubits 1 and 2 to the state of 
qubits $N$ and $N-1$ at the same transmission time $t^*$
(see also \cite{yao2013quantum} for a different, time dependent protocol). 
As the spin chain has to swap only the states of the four boundary qubits,
irrespective of what 
happens to the bulk qubits, it turns out that the full engineering of 
the chain is, again, not required. 
We first use a minimal scheme, namely we use a setup 
which does not require further experimental control in addition to that 
needed for the encoding and decoding procedures.
We show that the fidelity of transmission exceeds $89\%$ in the infinite chain
limit. Moreover, by introducing a further parameter we can achieve a 
transmission fidelity higher than $99\%$ for chains where $N\lesssim 500$.

The encoding/decoding procedure illustrated in Section~\ref{s.2fid} avoids
the destructive interference between the state to be transferred and the
excitations inside the bulk. The price to be paid is the full control of the
first two and last two qubits of the 
chain, as illustrated in Fig.~\ref{f.CD}. We 
argue accordingly that no more complications are introduced in requiring 
that the couplings $j_1=j_{N-1}$ and $j_2 = j_{N-2}$ can be switched on
and set to a particularly tuned value, while the rest of the qubits are
permanently coupled with a homogeneous interaction strength $j$
($j=1$ for convenience).
This model with two parameters has already been introduced in 
\cite{apollaro201299} with a different aim, i.e. 
for improving the transmission capability of an XX spin chain, 
without any encodings. Indeed, 
we have shown in Section~\ref{s.xx} that, with a single parameter, the two
competing effects for obtaining coherent ballistic dynamics cannot be 
optimized simultaneously and one has to choose a compromise. On the other
hand, in \cite{apollaro201299} it has been shown that using two parameters
these two constraints can be satisfied independently. 
Due to the lack of the encoding/decoding algorithm, 
that scheme is suitable only for particular initial states of $\tilde\Gamma$, 
and optimizes just the transmission from site 1 to site $N$, i.e.
from the sender to the receiver.
The figure of merit for the transmission quality is given by 
Eq.~\eqref{e.fidelity}, with $v_1(t)\equiv 0$ while $u_1(t)$ is given
by \eqref{e.wave}. Although
the analytic expressions for
$\varrho(q)$ and $\omega_q$ are slightly complicated because of $j_1$ and the 
further parameter $j_2$,  one can analytically prove that 
$F>99\%$ for $N\to\infty$. The resulting fidelity of state transmission 
is therefore almost perfect for every $N$.

\begin{table}
  \centering
  \caption{Transmission fidelity \eqref{e.fidelity} at the transmission time
  $t^*$, for different length $N$, and with the optimal couplings. 
  The left table
  refers to the cases discussed in \cite{apollaro201299}, where
  $j_1$ and $j_2$ are set to their optimal values.  
  In the right table on the other hand one also optimizes $j_3$ allowing a 
  higher value of $u_2(t^*)$ and accordingly a higher fidelity.  }
  \label{t.fid}       
  \begin{tabular}{ccccc}
    \hline\noalign{\smallskip}
    $N$ & $t^*$ & $j_1$ & $j_2 $ & $F_{\rm e}(t^*)$ \\
    \hline\noalign{\smallskip}
    20&26.5&0.550&0.818&0.987\\
    30&37.7&0.497&0.781&0.979\\
    50&59.5&0.434&0.735&0.970\\
    70&80.9&0.397&0.706&0.964\\
    100&112.5&0.359&0.675&0.958\\
    150&164.7&0.320&0.641&0.951\\
    250&267.8&0.276&0.599&0.944\\
    350&370.2&0.250&0.572&0.939\\
    500&523.0&0.225&0.544&0.935\\
    \noalign{\smallskip}\hline
  \end{tabular}
  \hspace{3mm}
  \begin{tabular}{ccccccc}
    \hline\noalign{\smallskip}
    $t^*$ & $j_1$ & $j_2 $ & $j_3$ & $F_{\rm e}(t^*)$ & $u_1(t^*)$ & $u_2(t^*)$  \\
    \hline\noalign{\smallskip}
    27.8&0.503&0.709&0.880&0.998&0.999&0.994\\
    39.4&0.448&0.648&0.846&0.996&0.999&0.991\\
    61.8&0.386&0.575&0.803&0.995&0.998&0.987\\
    83.6&0.349&0.529&0.775&0.994&0.998&0.985\\
    115.8&0.313&0.483&0.744&0.993&0.998&0.983\\
    168.6&0.276&0.434&0.710&0.993&0.998&0.982\\
    272.7&0.236&0.378&0.667&0.992&0.998&0.980\\
    375.9&0.212&0.344&0.640&0.992&0.998&0.979\\
    529.6&0.189&0.310&0.611&0.991&0.998&0.978\\
        \noalign{\smallskip}\hline
  \end{tabular}
\end{table}

Here we use that scheme for studying quantum state transfer with the two-qubit
encoding protocol. In this case, Eq.\eqref{e.Fapprox} shows that the
efficient transmission of one fermion from site 1 to $N$ is not enough and also
the transmission from site 2 to site $N-1$ has to be optimal. Although
the minimal engineering scheme with the two parameters 
$j_1$ and $j_2$ derived in 
\cite{apollaro201299} has not been designed for this purpose, 
Table~\ref{t.fid} shows that the final transmission quality $F_{\rm e}(t^*)$ is 
very good. Thanks to the analytical results available 
\cite{banchi2013spectral,apollaro201299} one can estimate the value of 
$F_{\rm e}(t^*)$ for $N\to\infty$. Indeed 
\begin{align}
  u_1(t^*) &\approx \left|\sum_q \varrho(q) e^{i [N q - \omega_q t^*]}\right|
  \overset{N\to\infty}{\simeq} 0.987~,  
  & 
  u_2(t^*) &\approx \left|\sum_q \frac{\omega_q^2}{j_1^2}
  \varrho(q) e^{i [N q - \omega_q t^*]}\right|
  \overset{N\to\infty}{\simeq} 0.712 ~,
  \label{e.uapproxi}
\end{align}
from which one estimates $F_{\rm e}(t^*) \approx 89.7\%$ for $N\to\infty$. The
first result in the above formula has been obtained in \cite{apollaro201299}
while the second one has been obtained by using the identities known for 
tridiagonal matrices \cite{banchi2013spectral} and going to the continuum 
limit. The lower value of $u_2(t^*)$ can be improved by
numerically optimizing $j_1$ and $j_2$ in order to maximize both $u_1(t^*)$ 
and $u_2(t^*)$. However, the results obtained are not significantly better:
even if one tries to maximise $u_2(t^*)$ alone, 
one obtains $u_2(t^*) \approx 0.732$ in the thermodynamic limit.

\begin{figure}[t]
 \centering
  \resizebox{0.45\textwidth}{!}{%
    \includegraphics{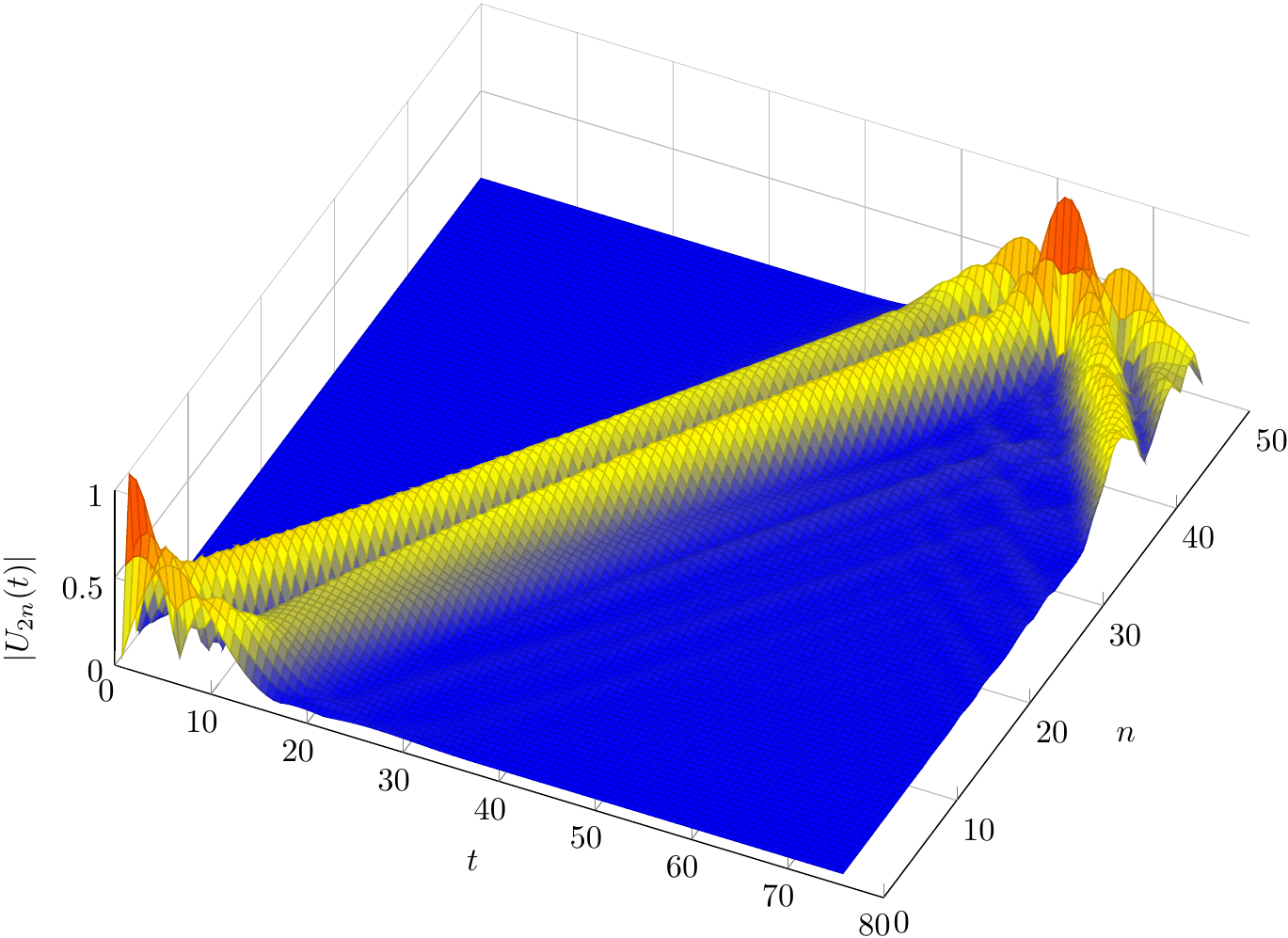}}
    \resizebox{0.45\textwidth}{!}{%
  \includegraphics{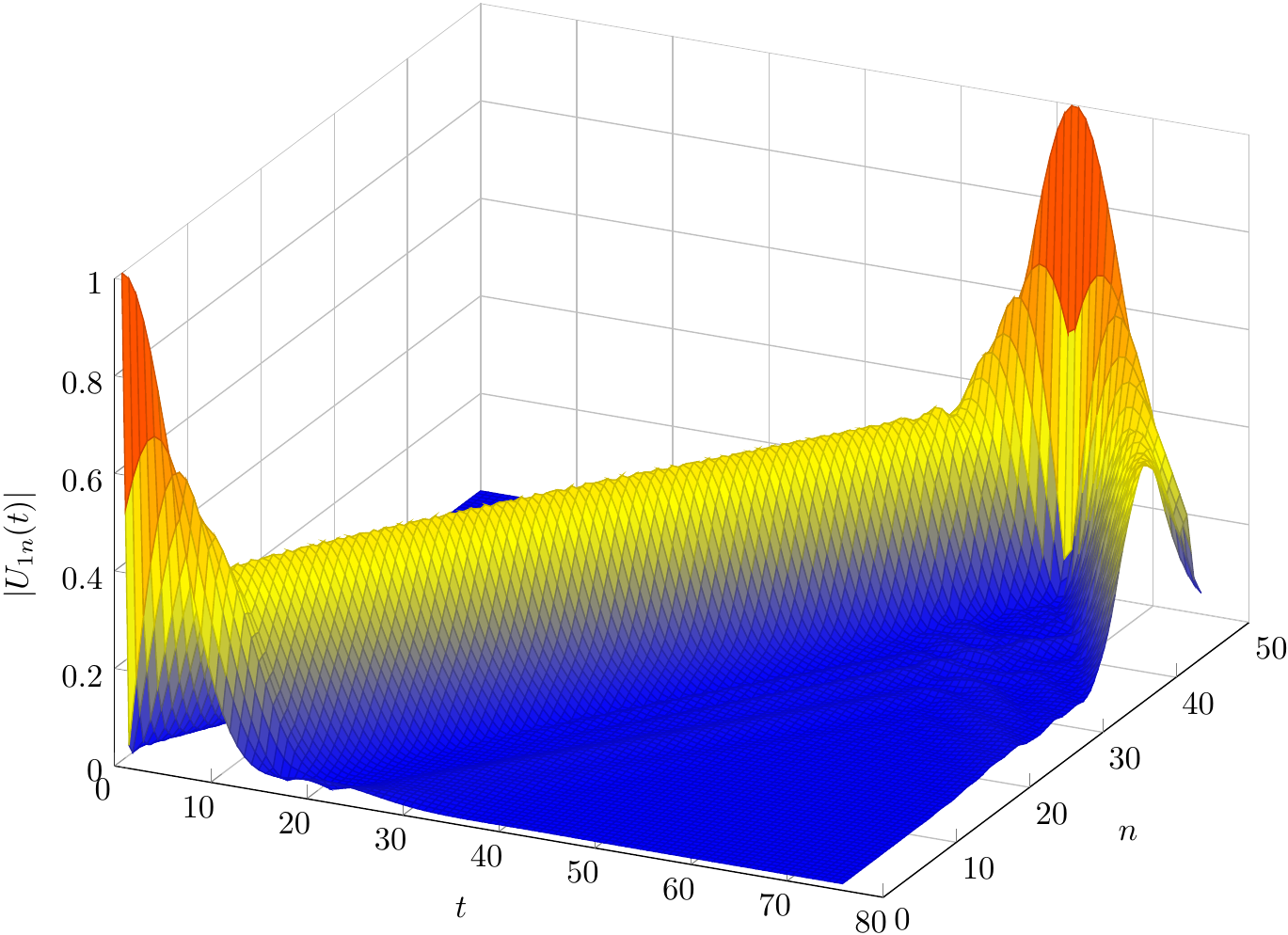}}
  \caption{Evolution operators $U_{1n}(t)$ and $U_{2n}(t)$ for different 
  position $n$ and time $t$. The wave-like motion displays the coherent
  excitation traveling from site 2 to site $N-1$ (left) and from site 1
  to site $N$ (right). An almost perfect reconstruction of the states occurs
  at the opposite sides when the optimal couplings are used. Here $N=50$ and
  the optimal couplings are shown in Table~\ref{t.fid}.}
 \label{f.3d}
\end{figure}

Almost perfect transmission for very long chains can be obtained by adding
a further parameter $j_3$ to be optimized. As shown in Table.~\ref{t.fid}, 
thanks to this further parameter one can obtain a fidelity higher than $99\%$
for chains as long as 500 qubits. The introduction of this further parameter
guides the creation of a coherent wave-packet both from site $1$ and from site
$2$. As shown in Fig.~\ref{f.3d} the excitation travelling
from site 2 generates two wave-packets. The first wave-packet goes 
towards site 1, where it is reflected from the boundary, and then goes towards
to the opposite end. The second wave-packet on the other hand goes directly 
towards the opposite boundary, where it is reflected. Thanks to the optimized
constructive interference, these two wave-packets come together in site 
$N-1$ at the transmission time with a probability $u_2(t^*)\approx 98\%$ 
for chains up to $N=500$. When this three-parameter optimization is used 
the transmission times are slightly longer, though they are 
still comparable with the previously obtained ballistic times 
(see Table~\ref{t.fid}).

Without the 
two-qubit encoding there is a temperature threshold, 
depending on the gap of the Hamiltonian, above which the transmission quality
is suppressed \cite{CamposVenutiBR2007}, though no significant alterations 
are expected in the dynamics of the $z$ component of the magnetization
\cite{cappellaro2011coherent}. In this section we have shown that with a 
two-qubit encoding one can reliably transfer quantum
information with minimal engineered chains even in the infinite temperature
regime.

\section{Towards experimental realizations}\label{s.exp}
The ballistic approach to state transfer using minimally engineered models
has been guided all along by the quest for experimental simplicity;
however,
there are some theoretical simplifications that must be overcome
in looking for possible implementations.  
Some general imperfections in the transfer scheme
may arise irrespective of the specific setup, such
as the gradual (non-sudden) switching of the coupling
between the sender (receiver) qubit and the bus, 
or the role of spurious interactions. 
In \cite{gate}
it has been shown that a linear switching of $j_1$ from
0 to $j_1^{\rm opt}$ does not alter significantly the transmission quality,
provided that the switching time is smaller than $1/j_1^{\rm opt}$.
Another experimentally relevant feature is that the overall ``reading time'' 
$t_{\rm R}$, 
i.e. the interval during which the state remains at the receiver site, 
is finite, as seen for instance in Fig.~\ref{f.3d}.  
In the optimal regime \cite{banchi2011long}, 
the transfer time scales with the system size as
$t^*\approx N + 2.3 N^{1/3}$ while the ``reading time'' increases with 
$N$ according to the asymptotic behaviour
$t_{\rm R} \approx 1.9 N^{1/3}$. As a result, the quantum channels based on
the optimal transfer scheme might be embedded also in a macroscopic setups.

Another detrimental effect can arise from the 
interaction between quasi-particles due to residual non-quadratic
terms in the Hamiltonian that cannot be completely screened out. 
These effects are investigated \cite{gate} with an XXZ
model. It is found that the introduction of 
a coupling $J^z$ along the $z$ direction
between adjacent spins evidently affects the quality of the channel. 
In fact, it weakly deteriorates the fidelity when
$|J^z|<0.2$, while the negative effects of this interaction 
are more pronounced for $J^z<-0.2$.
Moreover, the effect of static noise in the coupling strength 
has been investigated in \cite{zwick2011boundary,zwick2013optimal}.
A non-uniform deviation is introduced in the couplings of the internal 
chain,
$j_n\simeq 1+\delta_n$, $\delta_n\ll1$, while the optimal
parameters are supposed not to be affected by imperfections. 
It is shown that the quality of the 
transmission generated by the optimal couplings 
weakly deteriorates when $|\delta_n|\lesssim 0.05$. 
When the strength of the imperfections is stronger the performance of 
minimally engineered chains is comparable to that of fully engineered 
ones. If disorder cannot be avoided, the full engineering of the chain 
is therefore unnecessary, as minimally engineered models,
which are easier to implement in an experiment, 
have comparable transmission fidelities. 
On the other hand, imperfect tuning of the optimal parameters does not 
significantly alter the transmission quality.
For instance, in Fig.~\ref{f.jh} (see also \cite{banchi2011long})
one can see that there is an optimal region around the optimal values where
the fidelity remains very high.

In addition to the imperfections described above, whose effects depend on the
particular setup, there can also be other specific ones. As a matter of fact,
in order to use many-body dynamics for transferring quantum information, 
long coherence times are required, together with  
the ability of performing single-site
addressing and time-dependent measurements. 
An imperfect implementation of 
these requisites could introduce further error sources.
Understanding whether a currently established
experimental apparatus can be used as a coherent ballistic quantum data-bus 
is complicated, as one has to fathom if the various requisites 
might be somehow achieved in the near future.

In \cite{Yao2010,yao2012scalable} a chain of coupled nitrogen-vacancy (NV) 
defects in diamonds has been proposed to operate as a high-quality 
transmission wire, provided that $j_1\ll 1$. 
In \cite{ping2013practicality} a thorough analysis 
of the role of the typical decoherence times found in experiments has 
shown that a NV-wire can act as an entanglement transmission channel. 
It is there showed that non-perturbative couplings to the wire are 
favorable, so one is inclined to think that the techniques developed 
in this paper may
help the design of proper experimental setups based on NV centers.

Cold atom experiments 
\cite{bakr2009quantum,Brennen1999,mandel2003controlled,Greiner2002,sherson2010single,bakr2010probing,karski2010imprinting} 
are now  established \emph{quantum simulators}; 
they can effectively implement a spin chain and 
seem the most convenient for testing the predictions of this paper.
They have long coherence times, operating practically at zero temperature
and without decoherence.  For this reason, in \cite{gate} 
we have put forward a promising possible experimental realization using
cold atoms trapped in optical lattices and near field Fresnel trapping 
potentials. 

Nuclear magnetic resonance experiments are also promising candidates
for simulating quantum information transport
\cite{cappellaro2011coherent,cappellaro2007simulations,%
ramanathan2011experimental,ramanathan2011dynamics,%
ajoy2012algorithmic,rao2012entanglement}.
However, their highly mixed (high-temperature) initial state prevents the
possibility of using such a setup for transferring superpositions 
of quantum states without any other control on the system.
Coherent transmission of quantum superpositions can, in principle, be
performed using the two-qubit encoding and engineering proposed in 
Section~\ref{s.minimal2} provided that local gates acting on the 
boundaries can be implemented in the experimental setting.
Despite the difficulties in maintaining the phase coherence required
for transferring quantum superposition of states, some results of the
present paper might be observed even with currently available technology.
Indeed, the wave-packet generated by the two ``classical'' states
$\ket 0$ and $\ket 1$ survives, and it is not altered, 
even at finite temperature. The 
corresponding wave-like evolution \cite{apollaro201299} 
can then be observed by measuring the
magnetization along the $z$ direction $\mean{\sigma^z_n(t)}$.
The coherent magnetization dynamics allows the transmission of the two
{\it classical} states $\ket 0$ and $\ket 1$,
irrespective of what happens to their superposition, 
making the spin-wire still useful for classical information transfer.

\section{Concluding remarks}
\label{s.concl}

High-quality quantum state transmission between distant qubits 
can be obtained via the dynamics of a spin chain 
according to several different schemes. 
The one here reviewed relies on the optimization of the interactions between 
the boundaries 
(where the sender's and receiver's qubits are located) and the spin-chain
bus. Once such couplings are properly chosen,
the excitations with linear dispersion relation get to
rule the dynamics and a coherent ballistic transmission is consequently 
obtained.
The procedure does not require any further 
design either of the bus, or of its initial
state. When the approach is implemented with the spin-$1/2$ $XX$ and 
$XY$ models, the
problem can be mapped to the optimization of a Fermionic quantum walk 
in one dimension. 

Due to the
induced ballistic dynamics, the transfer time scale is considerably
shorter than in previous
works~\cite{Bose2007,WojcikLKGGB2005,PlastinaA2007,GualdiKMT2008,DoroninZ2010}.
It essentially depends on the group velocity of the elementary
excitations, which can be tuned by varying the global parameters of the
data-bus, namely the anisotropy $\gamma$ and the magnetic field $h$.
The optimal average fidelity, namely the figure of merit used for measuring
the transmission quality, in the optimal regime 
only slightly deteriorates as the length of the bus increases. 

A two qubit encoding protocol 
based on the optimal dynamics 
can also be designed such that 
the destructive effects of a (possible) thermal bath are overcome. 
The transfer quality is investigated using a specific 
figure of merit which takes into account 
the two-qubit encoding protocol.
It turns out that
the maximization of the transfer quality requires the optimization of the
dynamics so that the states of qubit 1 and 2 are transmitted to the states
of qubits $N$ and $N-1$. A minimal engineering which 
does not require further control on the system
is then introduced to achieve this goal.
The proposed protocol permits to
reliably perform quantum state transmission even at infinite temperature.

\section*{Acknowledgements}
The author thanks B.~Antonio, T.J.G.~Apollaro, A.~Bayat, S.~Bose, A.~Cuccoli,
R.~Vaia and P.~Verrucchi for interesting discussions and insightful comments.
The author is currently supported by the ERC grant PACOMANEDIA.

\appendix

\section{Diagonalization of quadratic Fermi Hamiltonian}\label{s.quaddiag}
The Hamiltonian \eqref{e.Hfermi} can be written in the form
\begin{align}
  H &= \frac12 \eta^\dagger\,S\,\eta+\frac12\Tr A, &
  S &= \begin{pmatrix}
    A & B \\-B&-A
  \end{pmatrix},
  \label{e.quadhamnotaz}
\end{align}
where $\eta_i = c_i$, $\eta_{i+N} = c_i^\dagger$ and $i=1,\dots,N$.
In \cite{bayat2011initializing,blaizot1986quantum} it is shown that 
the above Hamiltonian can be diagonalized via a canonical transformation
  and written in the form
\begin{align}
  H & = \frac12 \eta^\dagger\, 
  \begin{pmatrix} P & Q \\ Q & P \\ \end{pmatrix}^T\,
    \begin{pmatrix} \omega & 0 \\ 0 & -\omega \\   \end{pmatrix}\,
  \begin{pmatrix} P & Q \\ Q & P \\ \end{pmatrix}\,\eta +
    \frac12 \Tr A~ = 
   \frac12 \eta'^\dagger\, 
    \begin{pmatrix} \omega & 0 \\ 0 & -\omega \\   \end{pmatrix}\,
  \eta' +
    \frac12 \Tr A, 
    \label{e.quadhamdiag}
\end{align}
where the energies $\omega_k$ are diagonal and non-negative. 
That is, in terms of some diagonal
Fermi operators $b_k = \eta'_k$, $b_k^\dagger = \eta'_{k+N}$, $k=1,\dots N$
the Hamiltonian takes the form
$  H = \sum_k \omega_k\, b^\dagger_k\,b_k + \frac12\Tr(A-\omega)$. 
This transformation
diagonalizing the Hamiltonian can be obtained from a singular value 
decomposition \cite{LiebSM1961} and written as
\begin{align}
  A-B&=\Phi^T\,\omega\,\Psi~,& P &= (\Phi+\Psi)/2~,& Q&=(\Phi-\Psi)/2~.
  \label{e.diagXY}
\end{align}
where $\Psi$,$\Phi$ are orthogonal matrices and $\omega$ 
diagonal and non-negative.
The time evolution \eqref{ck_t} follows from the calculation of the
canonical transformation $e^{-i t S}$ and reads
\cite{bayat2011initializing} 
\begin{align}
  U(t) = P^T\, e^{\minus i \omega t}\, P  + Q^T\,e^{i \omega t}\, Q, &&
  V(t) = P^T\, e^{\minus i \omega t}\, Q  + Q^T\,e^{i \omega t}\, P.
  \label{e.evomatrix}
\end{align}


\subsection{Mirroring conditions for XY models}\label{s.xycondition}
A complication of the XY model is that the corresponding fermionic 
Hamiltonian does not conserve the number of particles. 
In this case the perfect mirroring condition reads $U(t^*)=e^{i\alpha}\; X$
and $V(t^*) = 0$, i.e. 
\begin{equation*}
  \begin{pmatrix} P^T & Q^T \\ Q^T & P^T \\ \end{pmatrix}\,
    \begin{pmatrix} e^{\minus i E t^*} & 0 \\ 0 & e^{i E t^*} \\ \end{pmatrix}\,
  \begin{pmatrix} P & Q \\ Q & P \\ \end{pmatrix} =
    \begin{pmatrix} e^{i\alpha}\; X & 0 \\ 0 & e^{-i\alpha}\; X\\
    \end{pmatrix}\equiv \hat X~.
\end{equation*}
As in the XX case, this means that the matrix on the r.h.s.~of the above 
equation can be diagonalized using the same matrices $P$ and $Q$, and 
that the energy-eigenvalues have to satisfy the condition \eqref{e.mirrorsym}.
Accordingly one must impose $[S,\hat X]=0$, i.e. that 
      $X\,A\,X=A$, $X\,B\,X =e^{2i\alpha}B$.
The matrix $A$ has to be persymmetric while there is still some freedom in
the symmetry properties of $B$. As $B$ is real 
the parameter $\alpha$, which in the $B=0$ case is a free parameter, 
has to be a multiple of $\frac\pi 2$: the matrix
$B$ thus has to be persymmetric, when for instance $\alpha=\frac\pi 2$ or 
anti-persymmetric, i.e. $\gamma_{N-n}=-\gamma_n$, when for instance 
$\alpha=\pi$. The physical origin of these constraints is still unclear,
but there is an argument supporting the persymmetric case $\alpha=
\frac\pi 2$.
Indeed, the XY model does not conserve the number of particles and the
phase $(-1)^{\frac{\bar n(\bar n-1)}2}$ in \eqref{e.stmirrored} is not
a constant of motion. However, the parity is conserved and 
$(-1)^{\frac{\bar n(\bar n-1)}2} e^{i\pi \bar n/2} = 
e^{i\pi \bar n^2/2}$ has a fixed value in each sector with constant parity. 
In the persymmetric case insofar, the dynamics effectively 
mirrors the state $\ket\psi$
without relative phases when the initial state has a definite parity.

We now generalize 
Lemma 2 of Ref.~\cite{kay2009review} for showing that, even in the
less stringent case of perfect transmission only between the boundary
qubits (irrespective of the bulk), 
the Hamiltonian has to satisfy some symmetries, and as a particular
case it can be persymmetric.  
Indeed, let us assume that 
\begin{equation*}
  \begin{pmatrix} P^T & Q^T \\ Q^T & P^T \\ \end{pmatrix}\,
    \begin{pmatrix} e^{\minus i E t^*} & 0 \\ 0 & e^{i E t^*} \\ \end{pmatrix}\,
  \begin{pmatrix} P & Q \\ Q & P \\ \end{pmatrix} 
    e_1  =  e^{i\alpha} e_N~,
    \label{e.bhobhobho}
\end{equation*}
being $e_i$ the vector with components $(e_i)_j = \delta_{ij}$. Then
it must hold 
\begin{align*}
  e^{-i\omega_kt^*} P_{k1} = e^{i\alpha} P_{kN}, &&
  e^{i\omega_kt^*} Q_{k1} = e^{i\alpha} Q_{kN}, && \forall k~.
\end{align*}
In particular, this reveals that $P_{k1}^2 = P_{kN}^2$,
$Q_{k1}^2 = Q_{kN}^2$, and $P_{k1}Q_{k1}=e^{2i\alpha}P_{kN}Q_{kN}$. 
Again the above constraints are satisfied only if $\alpha$ is a multiple
of $\pi/2$.
By rising the Hamiltonian matrix $S$ of 
\eqref{e.quadhamnotaz} to an integer power, $m$, we can relate 
\begin{align}
  e_1^TS^me_1 &= \sum_k \omega_k^m(P_{k1}^2+(-1)^mQ_{k1}^2) = e_N^TS^me_N~,
  &
  e_{N+1}^TS^{2m}e_1 &= 2\sum_k \omega_k^{2m}(P_{k1}Q_{k1}) = e^{i2\alpha} 
  e_{N+N}^TS^{2m}e_N ~.
\end{align}
For $m=1$, this gives that $h_1 = h_N$. For $m=2$, we find that 
$J_1^2-J_{N-1}^2 +\gamma_1^2-\gamma_{N-1}^2 = 0$ and
$J_1\gamma_1+e^{2i\alpha} J_{N-1}\gamma_{N-1} = 0$, 
i.e. that $j_1 = j_{N-1}$ and $\gamma_1
= -e^{2i\alpha}\gamma_{N-1}$. 
Each time that $m$ is increased by 1, new variables are
introduced on each side of the equation. Since these sides must be equals
the required symmetry properties follow. In particular we find that 
$B$ has to be persymmetric for $\alpha=\pi/2,3\pi/2$ and anti-persymmetric
for $\alpha=0,\pi$.

\subsection{Reflection symmetry: mirror symmetric XY models}\label{s.fermiJ}
The reflection symmetry, or mirror-symmetry,
occurs when the system is invariant under reflection,
i.e. when the Hamiltonian does not change by exchanging the sites 
$i$ and $N-i+1$, being $N$ the number of sites. 
Formally this means that $A_{i,j} = A_{N-j+1,N-i+1}$ and
$B_{i,j} = B_{N-j+1,N-i+1}$, i.e.,
\begin{align}
  X\,A\,X = A^T = A, && X\,B\,X=B^T=-B,
  \label{e.ABpersymm}
\end{align}
being $X$ the exchange matrix defined in \eqref{e.reflec}.
Matrices $P$ 
satisfying the condition $X\,P\,X=P^T$ are also called persymmetric. 
One important property of persymmetric matrices is that $P\,X$ and $X\,P$ 
are symmetric, and thus can be diagonalized by standard eigenvalue 
decomposition. Let 
\begin{equation*}
  (A-B)\,X = W^T\,\Omega\,W~,
\end{equation*}
be the eigenvalue decomposition of $(A-B)X$.
Eq~.\eqref{e.diagXY} is obtained by setting $W=\Phi$, $\omega=|\Omega|$
and $\Psi = s\,W\,X$, where $s=\sign\Omega$. 
Moreover, Eq.~\eqref{e.bhobhobho} for $\alpha = \pi/2$ impose that 
\begin{equation*}
  \begin{pmatrix} P & Q \\ Q & P \\ \end{pmatrix}\,
    \begin{pmatrix} X & 0 \\ 0 & -X \\   \end{pmatrix}\,
  \begin{pmatrix} P & Q \\ Q & P \\ \end{pmatrix}^\dagger
    = 
    \begin{pmatrix} x^+ & 0 \\ 0 & x^- \\   \end{pmatrix}~,
\end{equation*}
is diagonal, with diagonal $x^+$ and $x^-$. Explicit calculations show that
if $\omega$ is ordered in increasing order, then 
$x^+=-x^-=s$. 

%
%
%

%

\end{document}